\documentclass[reprint,showpacs,preprintnumbers,amsmath,amssymb]{revtex4-2}

\usepackage{upgreek}
\usepackage{graphicx}
\usepackage{amsmath}
\usepackage{amssymb}
\usepackage{physics}
\usepackage{comment}

\usepackage{upgreek}

\DeclareMathAlphabet{\mathcalligra}{T1}{calligra}{m}{n}

\usepackage{graphicx}
\usepackage{amsmath}
\usepackage{amssymb}
\usepackage{bbold}

\newcommand{\beq}{\begin{equation}}
\newcommand{\eeq}{\end{equation}}
\newcommand{\bdis}{\begin{displaymath}}
\newcommand{\edis}{\end{displaymath}}
\newcommand{\bea}{\begin{eqnarray}}
\newcommand{\eea}{\end{eqnarray}}
\newcommand{\barr}{\begin{array}}
\newcommand{\earr}{\end{array}}
\newcommand{\bfig}{\begin{figure}[!]}
\newcommand{\efig}{\end{figure}}

\begin{document}


\title{The spontaneous Nernst coefficient of ferromagnets from the interplay of \\ electron scattering and Berry curvature}
\author{Vittorio Basso}
\email{v.basso@inrim.it}
\author{Adriano Di Pietro and Alessandro Sola}
\affiliation{Istituto Nazionale di Ricerca Metrologica, Strada delle Cacce 91, 10135, Torino, Italy}

\date{\today}

\begin{abstract}
We employ the Boltzmann transport approach to derive the spontaneous Nernst coefficient for ferromagnetic metals, explicitly treating the transverse current density due to Berry curvature as a Fermi surface property. We find that the spontaneous Nernst coefficient is proportional to the inverse of the scattering time constant, implying that efficient spontaneous Nernst materials should exhibit relatively strong scattering, a stark contrast to ordinary Nernst materials. Furthermore, we establish a direct connection between the strength and sign of the spontaneous Nernst coefficient and the itinerant contribution to orbital angular momentum density arising from the Bloch bands. Finally we construct a rigid two-bands model to evaluate the thermoelectric coefficients by which we find a good agreement with the signs and orders of magnitude of the experimental coefficients of magnetic 3$d$ transition metal ferromagnets. We finally propose some practical recipes for maximizing the spontaneous Nernst effect through electronic band structure tailoring.
\end{abstract}

\maketitle

\section{Introduction}

The spontaneous or anomalous Nernst effect of metallic ferromagnets is currently an active topic of research with possible applications to the design of heat harvesters, heat flux sensors and active cooling devices \cite{Mizuguchi-2019,Uchida-2021,Elahi-2023,Adachi-2025}. Among the relevant materials we find MnBi \cite{He-2021,Sola-2023}, Fe$_3$Ga \cite{Sakai-2020,Stejskal-2023}, Ni$_2$MnGa \cite{De-2021} and Co$_2$MnGa \cite{Sakai-2018,Guin-2019}, and transition metal - rare earths compounds (Nd$_2$Fe$_{14}$B, SmCo$_5$, Sm$_{2}$Co$_{17}$) \cite{Miura-2019}. In these systems one has the coexistence of: i) relatively strong Nernst effects and ii) hard magnetic properties, able to permit to the device to work at the remanence without the need of an additional applied magnetic field. Unfortunately the figure of merit for the spontaneous Nernst effect is still orders of magnitude below the ordinary Nernst one ($z_T \sim 1$) and an effort in terms of search of new compounds or material optimization is therefore needed \cite{Li-2025}. However the direction to follow in order to optimize the active materials are not clear yet.

The spontaneous Nernst effect of ferromagnets shares its microscopic origin with the spontaneous Hall effect \cite{Xiao-2006}. This spontaneous contribution, typical of ferromagnets, is often called anomalous because it is not due to the Lorentz force like the ordinary one. The striking difference between spontaneous and ordinary effects is in the different dependence on electron scattering. While for the ordinary Hall effect the transverse electrical resistivity $\varrho_{H,xy}$ is an intrinsic, scattering independent, parameter, for the spontaneous effect it is the transverse electrical conductivity $\sigma_{s,xy}$ which is intrinsic and scattering independent \cite{Nagaosa-2010}. The origin of the spontaneous effect has been debated for decades \cite{Nagaosa-2010, Hurd-2012, Chien-2013}. Traditionally it has been interpreted in terms of different contributions: intrinsic \cite{Karplus-1954}, side jump \cite{Berger-1970} and skew scattering \cite{Smit-1955, Smit-1958}, however their relative importance remained elusive. The easiest contribution to disentangle is the skew scattering because it depends on $\tau_c$. If we limit to metals characterized by diffusive transport with relatively short mean free path (i.e. with conductivities $\sigma_e < 10^{8} \, \Omega^{-1}$m$^{-1}$) it is essentially suppressed \cite{Miyasato-2007, Nagaosa-2010, Weischenberg-2013}. The other two effects, intrinsic and side jump, are difficult to disentangle because both are scattering independent. Recently it has been clarified that the spontaneous transverse conductivity of ferromagnets, together with the spin Hall effect of metals and others transverse effects \cite{Sinova-2015}, are related to the Berry curvature of Bloch bands, a property accessible to first principle band structure calculation methods. Regarding the traditional subdivision into intrinsic and side jump mechanisms, it has been shown that both are related to the Berry curvature and independent of disorder \cite{Kovalev-2010, Weischenberg-2011}. Their relative relevance is then probably related to the transport model used. The modifications needed to the Boltzmann transport picture in order to account for the side jump have been described in Ref.\cite{Sinitsyn-2008}. By using the Kubo picture, Ref.\cite{Sushkov-2013} found that the side jump is irrelevant with respect to the other two while Ref.\cite{Weischenberg-2013} found that it is only a minor contribution at least for transition metal. The results obtained by computing the anomalous Hall conductivity $\sigma_{s,xy}$ by the tight binding \cite{Kontani-2007, Mitscherling-2020} or by the spin density functional theory (DFT) \cite{Yao-2004,Wang-2006,Yates-2007} showed good agreement with the experiments. A similar approach has been applied to compute the anomalous Nernst conductivity $\alpha_{s,xy}$, which is related to the energy dependence of $\sigma_{s,xy}$ \cite{Weischenberg-2013, Sawahata-2023, Stejskal-2023, Chiba-2024}, also showing good agreement with experiments \cite{Miyasato-2007, Pu-2008}. However this success has not led to find materials with an optimized Nernst effect yet. 

In the most general case (independently of the ordinary or spontaneous origin) the figure of merit, $z_T$, of transverse thermoelectrics is not related to Nernst conductivity $\alpha_{xy}$, but rather to the Nernst coefficient $\varepsilon_{N}$ given by

\beq
\varepsilon_{N} = \frac{\sigma_{xy} \alpha_{e}- \sigma_{e}\alpha_{xy}}{\sigma_{e}^2} \, ,
\label{EQ:Nerns_def0}
\eeq

\noindent where $\alpha_{e}$ is the longitudinal thermoelectric conductivity and $\sigma_{e}$ is the longitudinal electric conductivity \cite{Goldsmid-2016}. The Nernst coefficient $\varepsilon_{N}$ is the quantity actually measured in experiments and the studies aimed at the verification of the fundamental theories on the transverse thermoelectric conductivity $\alpha_{xy}$, have to derive its value from the measurement of the other four coefficients of Eq.(\ref{EQ:Nerns_def0}) \cite{Nakayama-2019, Sumida-2020}. Considering the efficiency of transverse thermoelectrics, the point is that the Nernst coefficient $\varepsilon_{N}$ is not scattering independent and in order to find its connection with material properties one has to study the scattering processes. Transport theories, including the effects of the Berry phase, have been discussed by Sinitsyn \cite{Sinitsyn-2008} and by Onoda et al. \cite{Onoda-2008} but the Nernst coefficient $\varepsilon_{N}$ and the figure of merit were not explicitly worked out. Recently Wang et al. \cite{Wang-2022} have used a heat-bath approach to derive the transverse thermal transport in presence of inelastic scattering but they have mainly focused their approach to the determine the anomalies on the Wiedemann-Franz ratio.

In the present study we address the problem of deriving the spontaneous Nernst coefficient of ferromagnetic metals $\varepsilon_{sN}$ by using the Boltzmann transport approach \cite{Wilson-1953, Solyom-2007}. To our aims we then limit, in the present work, to consider the standard Boltzmann transport picture in which the transverse effects arise from the spontaneous velocity term as a direct consequence of the Berry curvature of the Bloch bands \cite{Nagaosa-2010}. In this way we may loose some detailed effect due to specific scattering sources giving rise side jump and skew scattering effects, but we are able to grasp the main physics which is lying behind the transverse electron transport in ferromagnets. The central steps in our derivation is to write the transverse current density due to the Berry curvature of the bands as a Fermi surface property, a fact already noted by several authors \cite{Haldane-2004, Xiao-2006, Wang-2007} but not exploited to derive an expression for $\varepsilon_{sN}$. As a first result we find that the spontaneous Nernst coefficient $\varepsilon_{sN}$ is proportional to the inverse of the scattering time constant $\tau_c$. This fact is seen by using the Mott's expression for the thermoelectric conductivity 

\beq
\alpha_i = - \left(\frac{k_B}{e}\right) \frac{\pi^2}{3} k_BT \frac{\partial \sigma_i}{\partial \epsilon}\, ,
\label{EQ:alphadef}
\eeq

\noindent which is valid either for longitudinal ($i=e$) or for transverse ($i=xy$) effects and relates them to the energy dependence of the corresponding electrical conductivities. This relation immediately permits to write

\beq
\varepsilon_{l,N} = \left(\frac{k_B}{e}\right) \frac{\pi^2}{3} k_BT \frac{\partial}{\partial \epsilon}  \left( \frac{\sigma_{l,xy}}{\sigma_{e}}\right)\, ,
\label{EQ:epsNfirst}
\eeq

\noindent where $l=H$ is for the ordinary effect and $l=s$ is for the spontaneous effect. $\sigma_{e}$ is proportional to the scattering time constant $\tau_c$ while, the spontaneous effect, $\sigma_{s,xy}$ is intrinsic. Therefore efficient spontaneous Nernst materials should be found among magnetic materials with relatively strong scattering. This result contrasts with what is found for the ordinary Nernst effect (under magnetic field) for which efficient ordinary Nernst materials have to be found among materials with reduced scattering. Indeed, in that case, the Hall conductivity $\sigma_{H,xy}$ is proportional to $\tau_c^2$ and the Nernst coefficient results to be proportional to $\tau_c$ \cite{Goldsmid-2016}. 

As a second result we draw the connection between the spontaneous Nernst coefficient $\varepsilon_{sN}$ and the properties of the Bloch bands giving rise to $\sigma_{s,xy}$ through the Berry curvature. We are able to show that the spontaneous transverse conductivity is directly related to the density of orbital angular momentum due to the Bloch bands $\mathbf{l}_b$. The presence of an orbital angular momentum for Bloch electrons is well known in ferromagnets because it gives a small additional contribution to the magnetization (in addition to the spin) \cite{Coey-2010} and is the source of magneto-crystalline anisotropy \cite{Brooks-1940}. One part of the orbital angular momentum, $\mathbf{l}_a$, is formed at the atomic sites, but another part, $\mathbf{l}_b$, is due to the Berry curvature of the Bloch bands and is itinerant in character \cite{Xiao-2007,Ceresoli-2010}. We discuss a reasonable assumption, that we call the third Hund's rule for Bloch states, to understand the placement of the large $\mathbf{l}_a$ and $\mathbf{l}_b$ states inside energy bands of ferromagnets and we justify our assumptions by using a simplified two bands tight binging model. By making a rigid band model for 3$d$ ferromagnets we derive a recipe for the possible maximization of the Nernst effect by electronic band structure tailoring.

The paper is organized as follows. In section II we introduce the phenomenology of the transverse thermoelectric effects (called also thermomagnetic effects), in section III we discuss the Boltzmann transport approach in presence of the Berry curvature of the Bloch states and we demonstrate the Fermi surface properties of the spontaneous transverse kinetic coefficients. In section IV we show the dependence of the thermoelectric and thermomagnetic coefficients and of the figure of merit on the scattering time constant and in sections V we discuss the sign and the order of magnitude of the Nernst effect in transition metal ferromagnets. Finally in section VI we try to give predictions for the optimization of the Nernst effect within the series of 3$d$ ferromagnets and discuss the possible extensions to other magnetic metals.  

\section{Thermomagnetic effects}
\label{SECT:thermomagnetic}

In presence of a magnetic field or of a spontaneous magnetization we have the appearance of transverse transport effects called thermomagnetic effects. With the magnetic field and the magnetization along the $z$ axis we can describe the thermoelectric effects in the plane $(x,y)$. We get therefore a matrix relating the two in plane electrical current densities, $j_{e,x}$ and $j_{e,y}$, and the two heat current densities, $j_{q,x}$ and $j_{q,y}$, to the gradients of the electrochemical potential $\mu_e$ and of the absolute temperature $T$. The number of independent kinetic coefficients is reduced because of the Onsager reciprocal relations \cite{Callen-1948, Callen-1985, Solyom-2007}. For an isotropic metal (anisotropic metals are reviewed in Refs.\cite{Hurd-1974,Livanov-1999}) we get

\begin{figure}[htb]
\centering
\includegraphics[width=8cm]{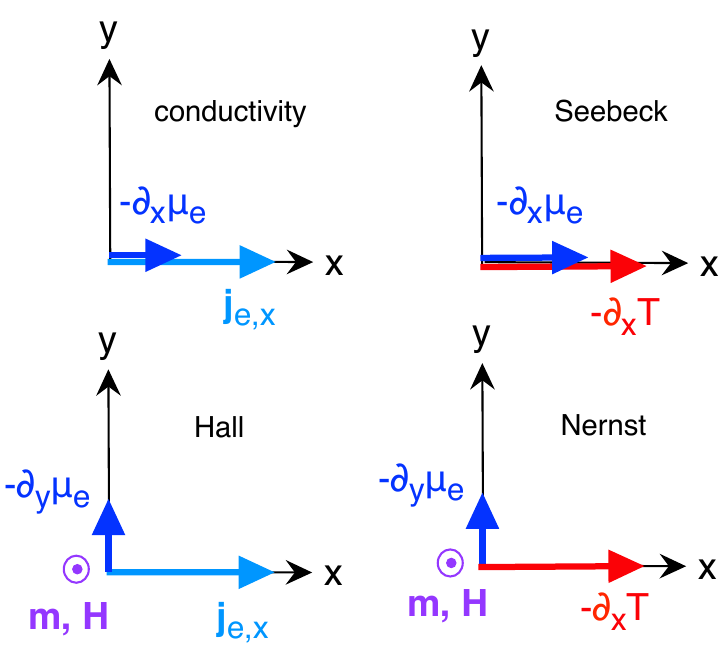}
\caption{Measurement conditions to determine the thermoelectric transport coefficients: the two longitudinal i.e. the conductivity $\sigma_e = j_{e,x}/(-\partial_y \mu_e)$ and the Seebeck thermopower $\varepsilon_e= \partial_x \mu_e / (-\partial_x T)$, and the two transverse, i.e. the Hall resistivity $\varrho_{xy} = (-\partial_y \mu_e)/j_{e,x} $ and the Nernst coefficient $\varepsilon_N =  (-\partial_y \mu_e)/(-\partial_x T)$.} \label{FIG:thermoelectric}
\end{figure}

\begin{widetext}
\beq
\left( 
\begin{array}{c}
j_{e,x} \\
j_{e,y} \\
j_{q,x} \\
j_{q,y} \\
\end{array}
\right)
=
\left( 
\begin{array}{cc}
\sigma_{e} 
\left(  \begin{array}{cc}
1 & -\theta_H \\
\theta_H & 1 \\ 
\end{array} \right) &
 \sigma_e \varepsilon_e
\left(  \begin{array}{cc}
1 & -\theta_N \\
\theta_N & 1 \\ 
\end{array} \right) \\
 \sigma_e \varepsilon_e T
\left(  \begin{array}{cc}
1 & -\theta_N \\
\theta_N & 1 \\ 
\end{array} \right)&
\kappa_e 
\left(  \begin{array}{cc}
1 & -\theta_R \\
\theta_R & 1 \\ 
\end{array} \right) \\
\end{array}
\right)
\left( 
\begin{array}{c}
-\partial_x \mu_e \\
-\partial_y \mu_e \\
-\partial_x T \\
-\partial_y T \\
\end{array}
\right) \, ,
\label{EQ:Matrix}
\eeq
\end{widetext}

\noindent where $\sigma_e$ is the electric conductivity, $\varepsilon_e$ is the thermopower and $\kappa_e$ is the thermal conductivity at constant electrochemical potential. To describe the transverse effects we have introduced the Hall angle $\theta_H$, the Nernst angle $\theta_{N}$ and the Righi-Leduc angle $\theta_{R}$. The angles $\theta_i$ depends on the magnetic field or on the magnetization and appear in the four ($x,y$) small-angle rotation submatrices of Eq.(\ref{EQ:Matrix}). The kinetic coefficients are the matrix elements relating currents and forces. They can be either computed by using statistical approaches for the electron transport such as the Boltzmann equation \cite{Wilson-1953, Solyom-2007} or determined experimentally. The measurable effects depend on the thermodynamic condition imposed (see Fig.\ref{FIG:thermoelectric}). The electric conductivity $\sigma_e = j_{e,x}/(-\partial_x \mu_e)$ (unit $\Omega^{-1}$m$^{-1}$) is defined under isothermal conditions $\partial_x T=0$. The thermopower $\varepsilon_e$ (unit VK$^{-1}$) is defined as $ \varepsilon_e = \partial_x \mu_e / (-\partial_x T)$ and measured under $j_{e,x}=0$. The longitudinal thermoelectric conductivity is given by their product $\alpha_{e} = \sigma_e\varepsilon_e$. The thermal conductivity, defined as $\kappa = j_{q,x} / (-\partial_x T)$ under $j_{e,x}=0$, includes all the contributions: electrons, phonons, magnons and so on $\kappa = \kappa^{(e)} + \kappa^{(p)}+\kappa^{(m)}+ ...$. The electronic contribution to the thermal conductivity, $\kappa^{(e)}$, is related to the kinetic coefficient $\kappa_e$ appearing in the matrix (Eq.(\ref{EQ:Matrix})) as $\kappa^{(e)} = \kappa_e - \sigma_{e} \varepsilon_e^2T$. To derive the expressions for the transverse effects we disregard any term in $\theta_i^2$ as small. Under isothermal conditions we get, with $j_{e,y}=0$, the transverse resistivity $\varrho_{xy} =  -\partial_y \mu_e/j_{e,x}$ (unit $\Omega$ m). The ordinary Hall effect is characterized by an intrinsic transverse resistivity $\varrho_{H,xy}$, proportional to the magnetic field. The spontaneous Hall effect of ferromagnets is characterized by an intrinsic transverse conductivity $\sigma_{s,xy}$. Therefore in presence of the two effects we get

\beq
\varrho_{xy} = \varrho_{H,xy} - \frac{\sigma_{s,xy}}{\sigma_e^2} \, .
\label{EQ:Transv_res}
\eeq

\noindent The isothermal Nernst coefficient or thermomagnetic power (unit VK$^{-1}$) is defined as $\varepsilon_{N} = -\partial_y \mu_e/(-\partial_x T)$ under $\partial_y T=0$ and $j_{e,x}=0$. The relation of $\varepsilon_{N}$ with the other coefficients is Eq.(\ref{EQ:Nerns_def0}), i.e.

\beq
\sigma_{e} \varepsilon_{N} = \sigma_{xy} \varepsilon_e - \alpha_{xy} \, ,
\label{EQ:Nerns_eq}
\eeq

\noindent where $\sigma_{xy} = \sigma_e\theta_{H}$ and the transverse thermoelectric conductivity is $\alpha_{xy} = \sigma_e\varepsilon_e\theta_{N}$. The efficiency of the thermoelectric devices is determined by the dimensionless figure of merit

\beq
z_T = \frac{ \sigma_{e} \varepsilon_i^2T}{\kappa}\, ,
\label{EQ:zT}
\eeq

\noindent which is valid for both longitudinal $i = e$ and transverse $i=N$ effects. Large $z_T$ corresponds to efficiency close to the Carnot value \cite{Goldsmid-2016}. To find paths for material optimization one has to derive the relation between $z_{T}$ and the microscopic parameter. In this paper we aim to clarify this connection for transverse spontaneous effects.

\section{Boltzmann transport for metals}

\subsection{Transport of Bloch electrons}

The theory of Bloch electrons in crystals is based on the description of the single electron wavefunction as $\psi_{n,\mathbf{k}}(\mathbf{r}) \propto \exp(i\mathbf{k}\cdot\mathbf{r}) u_{n,\mathbf{k}}(\mathbf{r})$ given by a plane wave multiplied by a lattice periodic function $u_{n,\mathbf{k}}(\mathbf{r})$. The quantum numbers of the Bloch electrons are therefore the band index $n$ and the wavenumber $\mathbf{k}$ defined in the Brillouin zone of the crystal. The equilibrium properties are fully described by the knowledge of the energy of the bands $\epsilon_n(\mathbf{k})$ and by the Fermi Dirac distribution function $f_0 = (1+\exp(\epsilon-\mu)/k_BT)^{-1}$ where $\mu$ is the Fermi energy. The non-equilibrium properties, in presence of a driving force, can be described by the change of the parameter $\mathbf{k}$ in time. This change persists until some scattering or collision process is able to bring the electron state back to the equilibrium distribution. The competition between these two effects is described by the Boltzmann transport equation \cite{Wilson-1953, Chambers-1990, Solyom-2007} giving the space and time evolution of the non equilibrium distribution function $f(\mathbf{k};\mathbf{r},t)$. The equation is

\beq
\frac{df}{dt} = \left.\frac{\partial f}{\partial t}\right|_{collisions}\, .
\label{EQ:Boltzmann}
\eeq

\noindent The collision term at the r.h.s of Eq.(\ref{EQ:Boltzmann}) can be evaluated, for example, in the relaxation time approximation

\beq
\left.\frac{\partial f}{\partial t}\right|_{collisions} = - \frac{f-f_0}{\tau_c} \, ,
\eeq

\noindent where $\tau_c$ is the relaxation time describing all the collision effects (see Ref.\cite{Pikulin-2011} for possible extensions beyond the relaxation time approximation). The effective force driving the system out of equilibrium is hidden inside the total time derivative at the left hand side of Eq.(\ref{EQ:Boltzmann}). For stationary states, $\partial f/\partial t = 0$, and considering the Lorentz force $\mathbf{F} = e \partial_{\mathbf{r}} \phi_e - e \mathbf{v}\times \mathbf{B}$ ($\partial_{\mathbf{r}}$ is the gradient in real space) acting on the Bloch electrons, where $\phi_e$ is the electrostatic potential and $\mathbf{B}$ is the magnetic field, one gets, by standard methods \cite{Wilson-1953, Solyom-2007}, the solution

\beq
f - f_0 = \tau \mathbf{v} \cdot \left(\mathbf{F}_{eff} + e \tau_c \mathbf{B} \times \left(  \frac{1}{m^*} \right)\mathbf{F}_{eff} \right) \, \left( - \frac{\partial f_0}{\partial \epsilon}\right)\, ,
\label{EQ:B-solution}
\eeq

\noindent where $(1/m^*)$ is the inverse of the effective mass tensor defined as

\beq
\frac{1}{m^*_{ij}} = \frac{1}{\hslash^2} \frac{\partial^2 \epsilon}{\partial k_j \partial k_i }\, .
\eeq

\noindent The previous solution is valid for low magnetic fields i.e. with $e B \tau_c/m^*\ll1 $ . The effective force is

\beq
\mathbf{F}_{eff} = -e (- \partial_{\mathbf{r}} \mu_e) + (\epsilon - \mu)\left( - \frac{\partial_{\mathbf{r}} T}{T}\right)\, ,
\eeq

\noindent where $\mu_ e = \phi_e - \mu/e$ is the electrochemical potential. The change of the wavenumber is therefore expressed in terms of the effective force as

\beq
\hslash \frac{\partial \mathbf{k}}{\partial t} = \mathbf{F}_{eff} - e \mathbf{v} \times \mathbf{B}\, .
\label{EQ:dkdt}
\eeq

\noindent In writing the previous equations we have dropped the band index $n$ for compactness. 

\subsection{Velocity of Bloch electrons}

As a next step, in other to address the transport properties, one has to derive the Bloch electron velocity $\mathbf{v}$ given by the commutator of the position operator with the Hamiltonian $\hslash \mathbf{v} = i [ \mathcal{H}, \mathbf{r}]$. It was not realized until recently that in out-of-equilibrium dynamic conditions the velocity has two terms \cite{Xiao-2006,Nagaosa-2010}, because the position operator $\mathbf{r}$ acts as $i \partial_{\mathbf{k}}$ on both the coefficients of the wavepacket, giving rise the well known group velocity, $\mathbf{v}_g = (1/\hslash) \partial_{\mathbf{k}} \epsilon_n$ ($\partial_{\mathbf{k}}$ is the gradient in reciprocal space), and the lattice periodic functions, $u_{n,\mathbf{k}}(\mathbf{r})$, giving rise to the additional transverse velocity term, called spontaneous velocity. The result is 

\beq
\mathbf{v} = \mathbf{v}_g - \frac{\mathbf{\Omega}_n}{\hslash} \times \hslash \frac{\partial \mathbf{k}}{\partial t} \, ,
\label{EQ:totalvelocity}
\eeq

\noindent where we have recovered the band index $n$. We see that the additional term is proportional to the time derivative of the wavenumber $\mathbf{k}$, therefore it is nonzero only under dynamic conditions. The proportionality factor $\mathbf{\Omega}_n = \partial_{\mathbf{k}} \times \boldsymbol{r}_n$ is a gauge independent vector field called Berry curvature (with unit m$^2$), given by the curl of the Berry connection $\boldsymbol{r}_n$ (with unit m, often called $\boldsymbol{\mathcal{A}}_n$ in the literature) defined as

\beq
\boldsymbol{r}_{n} = i \bra{u_n} \ket{\partial_{\mathbf{k}} u_n} 
\label{EQ:Berryconnection}
\eeq

\noindent The connection $\boldsymbol{r}_{n}$ is a gauge dependent quantity, i.e. its value depends on the arbitrary choice of the phase factor multiplying the wavefunction $\ket{u_n}$. $\boldsymbol{r}_{n}$ has the unit of a length and it can be seen as an intrinsic displacement of the position of the Bloch electron. Being gauge dependent, its  consequences can be seen only in the physical effects that eliminate the arbitrary gauge choice, for example: in presence of electron trajectories running along closed paths in the $\mathbf{k}$-space, giving rise to the acquisition of a geometrical Berry phase, or in the case of the electron transport, involving directly the  Berry curvature. Geometrically, $\mathbf{\Omega}_n$ can be seen as the curvature of the Hilbert space of the periodic part of the Bloch states $\ket{u_n}$ as a function of their $\mathbf{k}$-space mapping \cite{Berry-1989,Kolodrubetz-2017}. If we compose the system of Eq.(\ref{EQ:totalvelocity}) and Eq.(\ref{EQ:dkdt}) we can derive the velocity $\mathbf{v}$ and $\partial \mathbf{k}/\partial t$ as a function of the group velocity and of the effective force

\begin{widetext}
\beq
\left(\begin{array}{c} \mathbf{v} \\  \partial \mathbf{k}/\partial t\\
\end{array}\right)
= \frac{1}{1+(e/\hslash) \mathbf{B} \cdot \mathbf{\Omega}_n  }
\left(\begin{array}{cc} 1 & - \mathbf{\Omega}_n \times \\ (e/\hslash) \mathbf{B}\times & 1\end{array}\right)
\left(\begin{array}{c} \mathbf{v}_g \\ \mathbf{F}_{eff}/\hslash\\ \end{array}\right)\, .
\eeq
\end{widetext}

\noindent Then with $|(e/\hslash) \mathbf{B} \cdot \mathbf{\Omega}_n| \ll 1$ we can approximate the velocity as

\beq
\mathbf{v} \simeq \mathbf{v}_g - \frac{\mathbf{\Omega}_n}{\hslash} \times \mathbf{F}_{eff}
\label{EQ:velocity}
\eeq

\subsection{Current densities}
\label{SUBSECT:Current densities}

The current density of the physical quantity $Q$ is given by the integral

\beq
\mathbf{j}_{Q} = \frac{1}{(2\pi)^3} \sum_{n} \int_{v_k} Q \mathbf{v} \, f d^3k
\label{EQ:current}
\eeq

\noindent where $v_k$ is the volume of the Brillouin zone, the sum is over all the bands and $f$ is the non equilibrium distribution function. For the electric current $\mathbf{j}_e$ we use the charge $Q = -e$ while for the heat current $\mathbf{j}_q$ we use $Q = (\epsilon-\mu)$. The velocity $\mathbf{v}$ has the two terms given by Eq.(\ref{EQ:velocity}). Then also the current density is the sum of two terms $\mathbf{j}_{Q} = \mathbf{j}_{Q,c} +\mathbf{j}_{Q,s}$, a conventional term $\mathbf{j}_{Q,c}$ and a spontaneous term $\mathbf{j}_{Q,s}$. The conventional term is obtained by taking the group velocity and considering that the isotropic part of the distribution (see Eq.(\ref{EQ:B-solution})), gives no contribution. What remains is

\beq
\mathbf{j}_{Q,c} = \frac{1}{(2\pi)^3} \sum_{n} \int_{v_k} Q \mathbf{v}_g \, (f-f_0) d^3k
\label{EQ:currentwithvg}
\eeq

\noindent With $k_BT$ much smaller than the typical energy band width the derivative of the Fermi distribution $-\partial f_0/\partial \epsilon$ of Eq.(\ref{EQ:B-solution}) is sharply peaked at the Fermi level $\epsilon=\mu$ and we can make use of the Sommerfeld expansion to transform Eq.(\ref{EQ:currentwithvg}) in a Fermi surface integral. By transforming  the integral over the Brillouin zone volume $v_k$ as an integral over the energy and an integral over the surface $\Sigma_k$ one writes

\beq
\int_{v_k} ... d^3k = \int  \left(  \int_{\Sigma_k} ...  \frac{1}{\hslash v_g} d^2k \right) d\epsilon \, ,
\label{EQ:Fermiint}
\eeq

\noindent where $v_g$ is the modulus of the group velocity. The density of states (DOS) $\mathcal{D}_N = \partial N/\partial \epsilon$

\beq
\mathcal{D}_N = \frac{1}{(2\pi)^3} \sum_n \int_{\Sigma_F}  \frac{1}{\hslash v_g} d^2k \, 
\label{EQ:Int_DOS_N}
\eeq

\noindent (unit J$^{-1}$m$^{-3}$, $N$ is the volume density of electrons) is a Fermi surface integral with the integrand of Eq.(\ref{EQ:Fermiint}) equal to one. When the integrand of Eq.(\ref{EQ:Fermiint}) depends on the energy, it must be expanded in powers of $ \zeta = \epsilon-\mu$. The result reduces to Fermi surface integrals in which the power zero contributes one, the power one does not contribute and the power two contributes as $(\pi^2/3)(k_BT)^2$. The simplest example is the electric conductivity tensor corresponding to power zero and given by the integral 

\beq
\sigma_{e,ij} = \left( \frac{e^2}{h} \right) \frac{1}{(2\pi)^2} \sum_{n} \int_{\Sigma_F} \frac{\tau_c v_{g,i} v_{g,j}}{v_g}  d^2k \, ,
\label{EQ:condetens}
\eeq

\noindent where $(e^2/h) \simeq 38.7\cdot 10^{-6} \, \Omega^{-1}$ is the conductance constant so that the remaining factor of Eq.(\ref{EQ:condetens}) has unit m$^{-1}$. The conductivity explicitly depends on the integrand, which is proportional to the time constant $\tau_c$ evaluated at the Fermi surface. 

The spontaneous term of the current density is obtained by taking the spontaneous velocity and considering that the anisotropic part of the distribution, Eq.(\ref{EQ:B-solution}), gives a second order contribution in the effective force that can be neglected. What remains is

\beq
j_{Q, s, i} = \epsilon_{ijk} \left(- \frac{1}{\hslash} \frac{1}{(2\pi)^3} \sum_{n} \int_{v_k} Q \, \Omega_{n,j} F_{eff,k} f_{0}\, d^3k\right)  \, ,
\label{EQ:currentwithvs}
\eeq

\noindent where $\epsilon_{ijk}$ is the totally antisymmetric Levi-Civita tensor. Again, with $k_BT$ small, we can use the Sommerfeld expansion of $f_0$ and show that Eq.(\ref{EQ:currentwithvs}) can be transformed into an integral up to the Fermi level (the occupied states) of the expansion in powers of $\zeta$ of the integrand. However, as the integrand contains $\Omega_{n,j}$, which is a curl in the $\mathbf{k}$ space, we can show that also for the spontaneous effects the current density can be written as Fermi surface integral \cite{Haldane-2004}. By making again the example of the conductivity we can write the constitutive equation in vector form as $\mathbf{j}_{e,s} = \boldsymbol{\sigma}_s \times (-\partial_{\mathbf{r}} \mu_e)$ by using the spontaneous transverse conductivity as the vector

\beq
\boldsymbol{\sigma}_s = - \left( \frac{e^2}{h} \right) \frac{1}{(2\pi)^2} \sum_{n} \int_{v_F} \boldsymbol{\Omega}_n d^3k\, .
\label{EQ:sigmas0}
\eeq

\begin{figure}[htb]
\centering
\includegraphics[width=8cm]{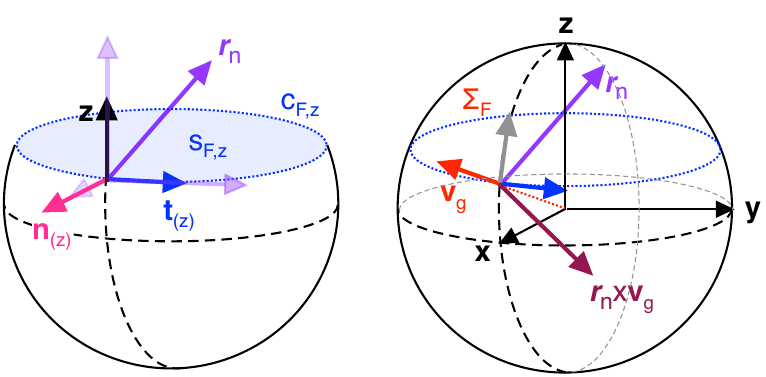}
\caption{Sketch of the Fermi surface $\Sigma_F$. Left. The line integral of Eq.(\ref{EQ:contourint}) corresponds to the projection of the displacement vector $\boldsymbol{r}_n$ onto the tangent unit vector $\mathbf{t}_{(z)}$ of the curve $c_{F,z}$. The result (along the $z$ axis) can be also written as $-(\boldsymbol{r}_n \cdot \mathbf{t}_{(z)}) \hat{\mathbf{z}} = \boldsymbol{r}_n \times \mathbf{n}_{(z)}$ where $\mathbf{n}_{(z)}$ is the normal to the $c_{F,z}$ curve. Right. By combining the three components we get the vector $\boldsymbol{r}_n \times \hat{\mathbf{v}}_g$.} \label{FIG:BZintegrals}
\end{figure}

\noindent Also in this case we choose to factor out the conductance constant $(e^2/h)$ and we are left with the remaining factor with unit m$^{-1}$. To transform the Fermi volume integral of Eq.(\ref{EQ:sigmas0}) into a Fermi surface one, we select one component of the curvature, say the $z$ component $\Omega_{n,z}$, and slice the Brillouin zone along the $(k_x,k_y)$ plane at given $k_z$. In this way it is possible to transform the volume integral into a surface integral over the $(k_x,k_y)$ slice of the surface $s_F(k_z)$ and an integral along $k_z$ (see Fig.\ref{FIG:BZintegrals} left). As the Berry curvature is the curl of the connection of Eq.(\ref{EQ:Berryconnection}) we write

\beq
- \int_{v_F} \Omega_{n,z} d^3k = - \int_{k_{z,min}}^{k_{z,max}} d k_z \int_{s_{F,z}} \, ( \partial_{\mathbf{k}} \times \boldsymbol{r}_n)_z \, d^2k \, ,
\label{EQ:intomega}
\eeq

\noindent then we apply the Stokes' theorem to the slice and get a contour integral in the $(k_x,k_y)$ plane along the closed path $c_F(k_z)$ surrounding the surface $s_F(k_z)$,

\beq
- \int_{s_{F,z}} \, (\partial_{\mathbf{k}} \times \boldsymbol{r}_n)_z \, d^2k = - \oint_{c_{F,z}} \, \boldsymbol{r}_n \cdot d\mathbf{k}
\label{EQ:contourint}
\eeq

\noindent The path $c_{F,z}$ is a Fermi contour at the energy $\epsilon=\mu$ and is subdividing the Brillouin zone torus $(k_x,k_y)$ (always at fixed $k_z$) into two regions: filled and empty states. The set of all the $c(k_z)$ paths can be parametrized by two variables: the energy $\epsilon$ and a phase parameter $\varphi$. The energy $\epsilon$ selects the curve while the phase parameter $\varphi$ runs along each curve. The integrand in Eq.(\ref{EQ:contourint}) corresponds to the projection of the displacement vector $\boldsymbol{r}_n$ along the tangent unit vector of the curve $c_{F,z}$ (see Fig.\ref{FIG:BZintegrals} left).  By adding the $k_z$ integral we obtain an integration over the Fermi surface $\Sigma_F$. By summing up the three components of the vector we finally obtain

\beq
\boldsymbol{\sigma}_s = \left( \frac{e^2}{h} \right) \frac{1}{(2\pi)^2}  \sum_{n} \int_{\Sigma_F} \boldsymbol{r}_n \times \hat{\mathbf{v}}_g \, d^2k \, ,
\label{EQ:sigmas}
\eeq

\noindent where $\hat{\mathbf{v}}_g$ is the unit vector of the group velocity which is normal to the Fermi surface. We have therefore transformed the Fermi volume integral of Eq.(\ref{EQ:sigmas0}) into the Fermi surface integral of Eq.(\ref{EQ:sigmas}).

Since the spontaneous transverse conductivity of Eq.(\ref{EQ:sigmas}) is physically observable and gauge independent, we should also be able to give a physical interpretation to its integrand. Indeed from the integrand we can define the angular momentum

\beq
\mathbf{l}_{b,n} = \boldsymbol{r}_n \times (m_e \mathbf{v}_g)\, ,
\eeq

\noindent where $m_e$ is the mass of the electron, and compute its volume density $\mathbf{L}_b$ by summing over the occupied states

\beq
\mathbf{L}_b = \frac{1}{2\pi} \int_{\epsilon_0}^{\mu} \left(   \frac{1}{(2\pi)^2} \sum_{n} \int_{\Sigma_F} \mathbf{l}_{b,n} \,\frac{1}{\hslash v_g}  d^2k \right) d\epsilon \, .
\label{EQ:Fermiint2}
\eeq

\noindent The spontaneous transverse conductivity results to be directly related to the density (as a function of the Fermi energy) $\boldsymbol{\mathcal{D}}_{b} = \left.\partial\mathbf{L}_b/\partial \epsilon\right|_{F}$ as

\beq
\boldsymbol{\sigma}_s = \left( \frac{e^2}{h} \right) \frac{h}{m_e} \boldsymbol{\mathcal{D}}_{b}
\label{EQ:sigmas2}
\eeq

\noindent The connection between the transverse conductivity and an orbital angular momentum is in fact expected. In the last few years it has been clarified that the orbital angular momentum of Bloch electrons is the sum of two terms \cite{Xiao-2007,Ceresoli-2010}. The first term, $\mathbf{l}_a$, is the atomic component, while the second, $\mathbf{l}_b$, is the band component. The atomic component $\mathbf{l}_a$ originates from the contribution of unquenched atomic orbitals and is spatially localized at the atomic sites \cite{Brooks-1940}. The second term, $\mathbf{l}_b$, is associated to the Bloch electron as a wavepacket and is caused by the Berry curvature vector $\boldsymbol{\Omega}_n$ \cite{Xiao-2007, Ceresoli-2010, Thonhauser-2011, Lopez-2012, Vanderbilt-2018, Aryasetiawan-2019, Dowinton-2022, Burgos-2024, Lee-2026}. While a magnetic field in the real space can cause curved electron trajectories, a Berry curvature in the reciprocal space gives rise to a non local contribution to the orbital angular momentum $\mathbf{l}_b$. The density $\boldsymbol{\mathcal{D}}_{b}$ appearing in Eq.(\ref{EQ:sigmas2}) is then labeled as the density of the itinerant angular momentum and it can be computed by either as a Fermi volume integral

\beq
\boldsymbol{\mathcal{D}}_{b}= - \frac{m_e}{h} \frac{1}{(2\pi)^2} \sum_{n} \int_{v_F} \boldsymbol{\Omega}_n d^3k\, . 
\label{EQ:Dbdefinitionwithvolume}
\eeq

\noindent or as a Fermi surface integral

\beq
\boldsymbol{\mathcal{D}}_{b} = \frac{1}{(2\pi)^3} \sum_{n} \int_{\Sigma_F} \mathbf{l}_{b,n} \,\frac{1}{\hslash v_g}  d^2k \, . 
\label{EQ:Dbdefinition}
\eeq

\noindent By making an integration by parts we can equivalently write $\mathbf{l}_{b,n} = (m_e/\hslash) (\epsilon_n-\mu) \boldsymbol{\Omega}_n$. This expression corresponds to the itinerant contribution to the orbital angular momentum derived within the theories of orbital magnetization in solids \cite{Thonhauser-2011}. Furthermore if we write Eq.(\ref{EQ:sigmas0}) with Eq.(\ref{EQ:Fermiint}) we get that the energy derivative of $\boldsymbol{\mathcal{D}}_{b}$,  is directly related to the Fermi surface integral of the Berry curvature

\beq
\frac{h}{m_e} \frac{\partial \boldsymbol{\mathcal{D}}_{b}}{\partial \epsilon}= - \frac{1}{(2\pi)^2} \sum_{n} \int_{\Sigma_k} \boldsymbol{\Omega}_n  \frac{1}{\hslash v_g} d^2k \, .
\label{EQ:Dbderivative}
\eeq

\noindent Eqs.(\ref{EQ:Dbdefinition}) and (\ref{EQ:Dbderivative}) will be of fundamental help in the definition of the spontaneous Nernst coefficient.

\section{Kinetic coefficients and the figure of merit}

The kinetic coefficients relating the current densities and the effective forces introduced in Section \ref{SECT:thermomagnetic} can be computed as integrals over the Fermi surface $\Sigma_F$. In the remaining of the paper the summation over the band index $n$ is understood to simplify the notation. In the determination of the kinetic coefficients it is essential to know the energy dependence of all the terms appearing in the integrand of Eq.(\ref{EQ:condetens}) and Eq.(\ref{EQ:sigmas2}). However the energy dependence is strongly different for the longitudinal and for the transverse effects. 

\subsection{Longitudinal effects}

The electric conductivity tensor is given by the integral of Eq.(\ref{EQ:condetens}) and depends explicitly on the time constant $\tau_c$. By placing the symmetry axis of the crystal along the reference frame we obtain a diagonal tensor and by assuming isotropic behavior we get the scalar conductivity

\beq
\sigma_{e} = \left( \frac{e^2}{h} \right) \frac{1}{(2\pi)^2}  \int_{\Sigma_F} \tau_c v_{g} \cos^2\theta \, d^2k \, ,
\label{EQ:sigmae}
\eeq

\noindent where $\theta$ is the angle between the velocity and the direction of conduction. A number of important observations can be made at this point. The first is that the collision time constant $\tau_c$ can be taken, to a good approximation, as independent of the direction $\mathbf{k}$, but dependent on the energy, then $\tau_c$ can be taken out of the integral. The remaining integration can be performed, in principle, by ab-initio DFT methods \cite{Gall-2016}. For the aims of the present work it is sufficient to observe that for transition metals the result is not strongly energy dependent. The conductivity can be written in a compact form by taking the average values at the Fermi surface of the integrand. We define the Fermi surface average of the quantity $Q(\mathbf{k})$ as

\beq
Q_F = \frac{1}{S_F} \frac{1}{(2\pi)^2} \int_{\Sigma_F}  Q(.)\, d^2k\, ,
\eeq

\noindent where $S_F = (2\pi)^{-2} \int_{\Sigma_F} d^2k$ (with unit m$^{-2}$) is the area of the Fermi surface and with the compact notation $Q(.)$ we mean that one has to distinguish between the argument kept fixed (i.e. the energy $\epsilon$) and the argument being integrated (i.e. the $\mathbf{k}$ values on the Fermi surface) as done in Eq. \ref{EQ:Fermiint}. Therefore the conductivity can be written as

\beq
\sigma_{e} = \left( \frac{e^2}{h}\right) \frac{1}{3}S_F v_F \tau_c\, ,
\label{EQ:sigmaeiso}
\eeq

\noindent where the factor (1/3) is the result of the average in three dimensions. The mean free path is the average distance travelled by an electron before scattering $\lambda = \tau_c v_F$.  The thermoelectric effects are obtained by taking the derivative with respect to the energy of the conductivity (Eq.(\ref{EQ:alphadef})). The thermoelectric power coefficient $\varepsilon_e$ is given by the Mott's formula \cite{Blatt-2012,Goldsmid-2016}

\beq
\varepsilon_e = - \left(\frac{k_B}{e}\right) \frac{\pi^2}{3} k_BT \frac{\partial \ln \sigma_e}{\partial \epsilon}
\label{EQ:Mott}
\eeq

\noindent The constant $(k_B/e) \simeq 86 \cdot 10^{-6}$ V/K is the scale of the thermoelectric effects and the remaining factor of the Mott's formula is dimensionless. The thermoelectric power depends on the energy variation of the conductivity, then on both the band structure and the type of scattering processes \cite{Witkoske-2017}. With the same method one gets the Wiedemann-Franz law for the thermal conductivity due to electrons

\beq
\frac{\kappa_{e}}{\sigma_e T} = \left( \frac{k_B}{e}\right)^2  \frac{\pi^2}{3}
\eeq

\noindent To have a large $z_T$ (Eq.(\ref{EQ:zT})) one has to first to minimize the thermal conductivity $\kappa = \kappa^{(e)} + \kappa^{(p)}$ and one possible strategy is to increase the lattice thermal resistivity. In the best conditions, with a negligible $\kappa^{(p)}$, we remain with $\kappa = \kappa^{(e)}$ and we get $z_T^{-1} = z_e^{-1} -1$ where

\beq
z_e = \frac{ \sigma_{e} \varepsilon_e^2T}{\kappa_e} = \frac{\pi^2}{3}\left( k_BT \frac{\partial \ln(\sigma_e)}{\partial \epsilon}\right)^2 \, ,
\eeq

\noindent is the electronic figure of merit $0< z_e <1$ independent of $\tau_c$. $z_e$ is typically maximized by selecting materials with a Fermi level lying very close to the beginning or to the end of an energy band as for example in doped semiconductors. 

As an example of an optimized longitudinal thermoelectric material we can take the Bi$_2$Te$_3$-type compounds. Many years of efforts have let to select doped $p$-type and $n$-type Bi$_2$Te$_3$ typically characterized by $\sigma_e \simeq 0.1 \cdot 10^6 \,  \Omega^{-1}$m$^{-1}$, $\varepsilon_{e} \simeq \pm 200 \cdot 10^{-6}$  VK$^{-1}$ and $\kappa \simeq 2$ Wm$^{-1}$K$^{-1}$ so that $z_T \simeq 0.6$ at $T \simeq 300$ K \cite{Goldsmid-2016}.

\subsection{Transverse effects under magnetic field}

The transverse thermoelectric effects in metals appear under an applied magnetic field \cite{Sondheimer-1948}. Take a magnetic field along $z$, $B_z$, and currents and forces in the perpendicular plane $(x,y)$. To simplify the description assume isotropic behavior in the plane. The Hall conductivity can be obtained by combining Eq.(\ref{EQ:currentwithvg}) and Eq.(\ref{EQ:B-solution}), resulting
 
\beq
\sigma_{H,xy} = \left( \frac{e^2}{h}\right) \frac{1}{(2\pi)^2}  \int_{\Sigma_F} \frac{\tau_c^2 e B_z}{m^*} v_g \cos^2\theta d^2k\, .
\label{EQ:sigmathetaH}
\eeq

\noindent Eq.(\ref{EQ:sigmathetaH}) can be written in a compact way by using the Fermi surface averages as

\beq
\sigma_{H,xy} = \left( \frac{e^2}{h}\right) \frac{1}{3} S_Fv_F \tau_c^2 \omega_c \, ,
\label{EQ:sigmathetaH2}
\eeq

\noindent where $\omega_c = e B_z/m^*$ is the cyclotron speed computed by using the Fermi surface averaged effective mass $m^*$. The Hall angle $\theta_{H} = \omega_c \tau_c$ is proportional to $\tau_c$ while the Hall resistivity $\varrho_{H,xy} = -\sigma_{H,xy}/\sigma_e^2$ results to be an intrinsic coefficient, independent of $\tau_c$. The Nernst coefficient is computed by Eq.(\ref{EQ:epsNfirst}) resulting in the Sondheimer's formula \cite{Sondheimer-1948, Delves-1965,Zebarjadi-2021} 

\beq
\varepsilon_{N} = \left(\frac{k_B}{e}\right) \frac{\pi^2}{3} k_BT \frac{\partial \theta_{H}}{\partial \epsilon} \, .
\label{EQ:MottHall}
\eeq

\noindent By passing to the Fermi surface average values we observe that the Nernst effect is due to the energy dependence of the effective mass $m^*$ and of the time constant $\tau_c$. The thermal Hall angle results to be equal to the Hall angle $\theta_R = \theta_H$. As the Nernst coefficient is proportional to $\tau_c$, and assuming that the electronic thermal conductivity is the dominant contribution, the whole $z_{T}$ of Eq.(\ref{EQ:zT}) rescales with $\tau_c^2$. Therefore efficient materials with the ordinary effect must have a large $\tau_c$, i.e. minimal scattering.

As an example of a material for ordinary Nernst refrigeration we can take the semimetal Bismuth. Good quality crystals have $\sigma_e \simeq 0.75 \cdot 10^6 \,  \Omega^{-1}$m$^{-1}$, $\varepsilon_{N} \simeq 20 \cdot 10^{-6}$  VK$^{-1}$ (polycrystalline at $B = 1$ T \cite{Sola-2025}) and $\kappa \simeq 7.5$ Wm$^{-1}$K$^{-1}$ \cite{Yim-1972}. At $T \simeq 300$ K we get only $z_T \simeq 0.012$, however at low temperature ($T \simeq 80$ K) the scattering with phonons is substantially reduced (and $\tau_c$ increases correspondingly) so that a much larger value $z_T \simeq 0.24$ is obtained. This makes the ordinary thermomagnetic effect a possible alternative to the longitudinal one only at low temperatures and with high quality materials \cite{Goldsmid-2016}. 

\subsection{Spontaneous transverse effects}

The spontaneous transverse conductivity is given by Eq.(\ref{EQ:sigmas2}). By taking only effects in the $x,y$ plane we get

\beq
\sigma_{s,xy} = \left( \frac{e^2}{h} \right) \frac{h}{m_e} \mathcal{D}_{b,z} \, ,
\label{EQ:sigmathetas}
\eeq

\noindent where $\mathcal{D}_{b,z} = \left.\partial L_{b,z}/\partial \epsilon\right|_{F}$ (unit s m$^{-3}$) is the density of the itinerant angular momentum at the Fermi level (Eq.(\ref{EQ:Dbdefinition})).  Another way to write the conductivity is to employ directly Eq.(\ref{EQ:sigmas}) and take the Fermi surface averages to get

\beq
\sigma_{s,xy} = \left( \frac{e^2}{h} \right) S_F r_b\, ,
\label{EQ:sigmathetas2}
\eeq

\noindent where $r_b$ (unit m) is the Fermi surface average of the intrinsic displacement $(\boldsymbol{r}_n \times \hat{\mathbf{v}}_g)_z$ from Eq.(\ref{EQ:Berryconnection}). We can write the spontaneous Hall angle $\theta_{sH} = \sigma_{s,xy}/\sigma_e$ as the ratio between the two lengths: intrinsic displacement $r_b$ and (one third of the) mean free path $\lambda$:

\beq
\theta_{sH} = \frac{r_b}{\lambda/3}\, .
\label{EQ:thetas}
\eeq

\noindent The spontaneous Nernst coefficient is computed by Eq.(\ref{EQ:Nerns_def0}) resulting in

\beq
\varepsilon_{sN} = \left(\frac{k_B}{e}\right) \frac{\pi^2}{3} k_BT \frac{\partial \theta_{sH}}{\partial \epsilon} \, .
\label{EQ:MottSpont}
\eeq

\noindent In this case we get the spontaneous Nernst coefficient depending upon the inverse of $\tau_c$, therefore the whole $z_{T}$ for the spontaneous Nernst effect rescales with $\tau_c^{-2}$ then making the increase of the scattering processes convenient for maximization of $z_{T}$. The previous relations are however valid for a single spin band. For metals with spin inversion symmetry they are all zero. For ferromagnets, we have to consider the two spin bands and write the expressions for conductivity and thermopower making specific assumptions on the band structure (see below, in Section \ref{SECTION:Model with two d orbitals}). 


As an example of a ferromagnet that can be employed as spontaneous transverse thermoelectric we take MnBi. At room temperature MnBi has a large uniaxial anisotropy and the material can be easily prepared with small grain size in order to develop hard magnet properties. Therefore a thermomagnetic device of MnBi (once magnetized) can work in absence of a magnetic field. MnBi has $\sigma_e \simeq 1.2 \cdot 10^6 \,  \Omega^{-1}$m$^{-1}$, $\varepsilon_{sN} \simeq -2 \cdot 10^{-6}$  VK$^{-1}$ and $\kappa \simeq 7$ Wm$^{-1}$K$^{-1}$ \cite{He-2021, He-2021b, Sola-2023} so that at $T \simeq 300$ K we get the extremely low value of $z_T \simeq 2.1 \cdot 10^{-4}$. Other candidate transverse thermoelectric materials have $z_T$ values in the same range: $z_T\simeq 3.2 \cdot 10^{-4}$ for Co$_2$MnGa \cite{Guin-2019} and $z_T\simeq 4.5 \cdot 10^{-4}$ for SmCo$_5$ \cite{Miura-2019}.  Then, in order to make the spontaneous thermomagnetic effect of interest for any type of application, one has to act on any possible source of $z_T$ improvement.

\section{Transition metal ferromagnets}

In this part of the paper we make use of the background obtained so far to find the line of improvement of $z_T$ of the spontaneous Nernst effect. A detailed derivation of the electron transport requires not only the detailed knowledge of its band structure but also of the scattering processes. Therefore here we limit to the case of transition metal ferromagnets, at temperatures well below the Curie point, in which it is possible to make a few simplifying assumptions on the band structure and on the scattering processes \cite{OHandley-2000}. This assumptions will provide compact expression for the transport coefficients $\sigma_{e}$, $\varepsilon_e$,  $\sigma_{s,xy}$ and $\alpha_{s,xy}$.

\begin{figure}[htb]
\centering
\includegraphics[width=6cm]{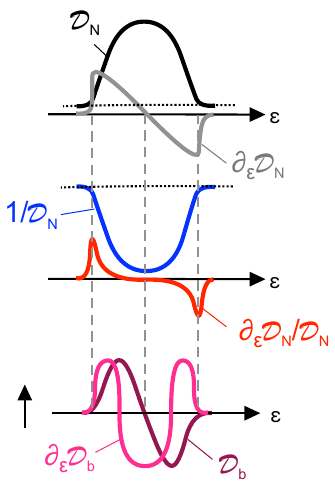}
\caption{Sketch of the density of states $\mathcal{D}_N$ and of the density of itinerant angular momentum $\mathcal{D}_b$ in the case of unquenched orbitals and for a single spin band. i) Top, density of states $\mathcal{D}_N$ and its energy derivative. ii) Center, for transition metals the conductivity of Eq.(\ref{EQ:condMott}) is proportional to the inverse of the density of states $1/\mathcal{D}_N$ and the thermopower is given by Eq.(\ref{EQ:thermopowerMott}). iii) Bottom, in the case of a majority band (spin negative) the density of itinerant angular momentum $\mathcal{D}_b$ is positive at low energy and negative at high energy.} \label{FIG:Bands1c}
\end{figure}

\subsection{The Fermi golden rule for scattering}

The conductivity, Eq.(\ref{EQ:sigmaeiso}), contains two main factors: the product $S_Fv_F$, evaluated at the Fermi surface, and time constant $\tau_{c}$. The product $S_Fv_F$ is accessible to ab-initio DFT methods \cite{Gall-2016}, however the main point is that only a small portion of the Fermi surface, with states characterized of low effective mass (light carriers), contributes significantly to the $S_Fv_F$, while most of the Bloch states, characterized by a much larger mass (heavy carriers), contributes very little. On the other hand, the scattering time constant $\tau_c$, must be determined by the Fermi golden rule i.e. the scattering rate $1/\tau_c$ is proportional to the density of states $\mathcal{D}_{N}$ at the Fermi level (unit J$^{-1}$m$^{-3}$), i.e. $1/\tau_{c} = f_c \mathcal{D}_{N}$, where $f_c$ is the scattering strength parameter (unit Jm$^{3}$s$^{-1}$) independent of the energy but dependent on temperature (see Appendix). These two simplifying assumptions have been extensively considered as a reasonable model for transition metals with 4$s$ and 3$d$ electrons \cite{Mott-1964, Garmroudi-2023}. The hybridization between $s$ and $d$ electrons produces a subset of $s$-type light electrons contributing to the product $S_Fv_F$ and another subset of $d$-type heavy electrons contributing greatly to the density of states. This second subset acts as a trap for the first mobile subset. The conductivity, Eq.(\ref{EQ:sigmaeiso}), results

\beq
\sigma_{e} = \left( \frac{e^2}{h} \right) \frac{1}{3} \frac{S_Fv_F}{f_c} \frac{1}{\mathcal{D}_{N}}\, .
\label{EQ:condMott}
\eeq

\noindent To compute the thermoelectric coefficient, Eq.(\ref{EQ:Mott}), we assume that the factor $S_Fv_F$ does not depend significantly on the energy, while the density of states $\mathcal{D}_{N}$ does. Then the main energy variation of the conductivity is contained in the $\tau_c$ factor. We get

\beq
\varepsilon_{e} =\left(\frac{k_B}{e}\right) \frac{\pi^2}{3} k_BT \frac{\partial \ln \mathcal{D}_{N}}{\partial \epsilon}\, ,
\label{EQ:thermopowerMott}
\eeq

\noindent where now, with respect to Eq.(\ref{EQ:Mott}), the sign is plus instead of minus. In non ferromagnetic or paramagnetic metals the density of states due to the $d$ bands is rich of peaks and valleys, however as a general rule, one expects, in a naive rigid band approach, that varying continuously the Fermi level from the beginning to the end of the $d$ band one goes from positive to negative thermoelectric power. This is indeed the trend which is observed experimentally \cite{Blatt-2012}. Close to the complete filling of the $d$ band one finds the constantan alloys (Ni$_{45}$Cu$_{55}$) with $\varepsilon_e \simeq - 45\cdot 10^{-6}$ VK$^{-1}$. The alloy Ni$_{x}$Au$_{1-x}$ with $x=0.37$ has been recently shown to display one the largest thermopower, $\varepsilon_e \simeq - 75 \cdot 10^{-6}$ VK$^{-1}$, among metals. A sketch of the DOS is shown in Fig.\ref{FIG:Bands1c} top and center.

\subsection{The Mott's two currents model}
\label{The_Mott_s_two_currents_model}

In the Mott's two currents model one considers the spin flip process as highly unlikely (an assumption which is reasonably correct for light transition metals, see Appendix) so that the two spin channels for transport can be considered as independent \cite{Mott-1945, Coey-2010}. We take $\uparrow$ for the majority electrons (those contributing to positive magnetization along $z$ for ferromagnets, i.e. with negative spin) and $\downarrow$ for the minority electrons. The conductivity is then given by the sum of the two independent spin contributions. We may assume that the product $S_Fv_F$ is approximately the same for the two spin bands. Then we write the conductivity as

\beq
\sigma_{e} = \left( \frac{e^2}{h} \right) \frac{1}{3} \frac{S_Fv_F}{f_c}\left(\frac{1}{\mathcal{D}_{N,\uparrow}} +\frac{1}{\mathcal{D}_{N,\downarrow}}  \right) \, ,
\label{EQ:sigmaeTMF}
\eeq

\noindent and the thermoelectric coefficient as

\beq
\varepsilon_{e} =\left(\frac{k_B}{e}\right) \frac{\pi^2}{3} k_BT \frac{\mathcal{D}_{N,\downarrow}   \partial_{\epsilon} \ln\mathcal{D}_{N,\uparrow}
 + \mathcal{D}_{N,\uparrow}   \partial_{\epsilon} \ln \mathcal{D}_{N,\downarrow}
 }{\mathcal{D}_{N,\uparrow} +\mathcal{D}_{N,\downarrow}}\, ,
\label{EQ:epsiloneTMF}
\eeq

\noindent where $\mathcal{D}_{N,i}$ with ($i = \uparrow, \downarrow$) is the density of states for each spin band.

\begin{figure}[htb]
\centering
\includegraphics[width=8.5cm]{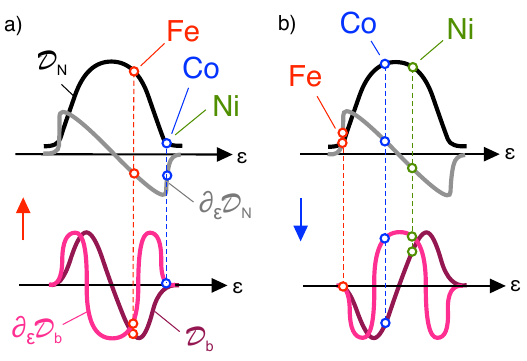}
\caption{The two spin bands of antibonding states of ferromagnets. Top: sketch of the density of states $\mathcal{D}_N$ and its energy derivative $\partial_{\epsilon}\mathcal{D}_N$. Bottom: sketch of the density of itinerant angular momentum $\mathcal{D}_b$ and its energy derivative $\partial_{\epsilon}\mathcal{D}_b$. a) Majority band. For Fe it is partially filled and the Fermi level lies in the region of negative orbital momentum $\mathcal{D}_b$. For Ni and Co it is completely filled (strong ferromagnets). b) Minority band. For Fe it is empty (weak ferromagnet) while for Ni and Co it is partially filled but the Fermi level lies in the region of negative orbital momentum $\mathcal{D}_b$ for Co and positive for Ni. } \label{FIG:Bands3dbis}
\end{figure}

\begin{table*}
\begin{center}
\begin{tabular}{|c|c|c|c|c|c|c|c|c|c|c|}
	\hline
& $\sigma_{e}$ & $\varepsilon_{sN}$ & $\sigma_{e}\varepsilon_{sN}$& $\sigma_{s,xy}$ & $\varepsilon_{e}$  & $\sigma_{s,xy}\varepsilon_{e}$ & $-\alpha_{s,xy}$ & $\kappa$ & $z_T$ \\
& [$\Omega^{-1} $m$^{-1}$] & [V K$^{-1}$] & [A m$^{-1}$K$^{-1}$]  & [$\Omega^{-1} $m$^{-1}$] & [V K$^{-1}$] & [A m$^{-1}$K$^{-1}$] & [A m$^{-1}$K$^{-1}$] & [WK$^{-1}$] & \\
& ($\cdot 10^{6}$) & ($\cdot 10^{-6}$) && ($\cdot 10^{6}$) & ($\cdot 10^{-6}$) & & & & ($\cdot 10^{-6}$) \\
	\hline
Fe  & 8.6$^{b}$ & $-$0.44$^{b}$ &$-$3.9& $-$0.103$^{d}$     & 14$^{a}$ &$-$1.4& $-$2.5 & 80 & 6.2\\
Co  & 9.7$^{c}$ & 0.39$^{c}$ &3.8& $-$0.020$^{e}$ & $-$32$^{a}$  &0.6& 3.2 & 70 & 6.3 \\
Ni   & 8.2$^{b}$  & 0.34$^{b}$  & 2.8 & 0.066$^{f}$ & $-$20$^{a}$  &$-$1.3& 4.1& 90 & 3.2 \\
	\hline
\end{tabular}
\end{center}
\caption{Experimental thermomagnetic transport coefficients of 3$d$ ferromagnets, Fe, Co, and Ni around room temperature, taken from the literature: thermoelectric coefficient $\varepsilon_{e}$ from Ref.\cite{Weischenberg-2013} (a), Nernst coefficient $\varepsilon_{sN}$ and the electric conductivity $\sigma_{e}$ from Ref.\cite{Weischenberg-2013} and specifically from Ref.\cite{Butler-1940}(b) (Fe, $T=313$ K and Ni, $T=313$ K, polycrystals), Ref.\cite{Hall-1925}(c) (Co, $T=320$ K); spontaneous Hall conductivity $\sigma_{s,xy}$ from Ref.\cite{Wang-2007} (the signs are adapted to the convention of Eq.(\ref{EQ:Transv_res})) and specifically from Ref.\cite{Dheer-1967}(d) (Fe, $T=298$ K), Ref.\cite{Volkenshtein-1960}(e) (Co, $T=293$ K) and Ref.\cite{Lavine-1961}(f) (Ni, $T=293$ K); thermal conductivity $\kappa$ from Ref.\cite{Stearns-1986}; $z_T$ is computed by Eq.(\ref{EQ:zT}).}
\label{TABLE:Spontaneous}
\end{table*}

\subsection{The third Hund's rule for Bloch states}
\label{SUBSECT:The third Hund's rule for Bloch states}

The spontaneous effects are given by the transverse conductivity of Eq.(\ref{EQ:sigmathetas}). We consider only the $z$ component and the two spin bands, writing

\beq
\sigma_{s,xy} = \left( \frac{e^2}{h} \right) \frac{h}{m_e} (\mathcal{D}_{b,\uparrow}+\mathcal{D}_{b,\downarrow}) \, .
\label{EQ:sigmasxyTMF}
\eeq

\noindent The Nernst coefficient is computed by Eq.(\ref{EQ:Nerns_eq}), i.e. $\sigma_{e} \varepsilon_{N} = \sigma_{s,xy} \varepsilon_e - \alpha_{s,xy}$, where

\beq
- \alpha_{s,xy} = \left(\frac{k_B}{e}\right) \frac{\pi^2}{3} k_BT \left( \frac{e^2}{h} \right) \frac{h}{m_e} \left(\frac{\partial \mathcal{D}_{b,\uparrow}}{\partial \epsilon}+ \frac{\partial \mathcal{D}_{b,\downarrow}}{\partial \epsilon}\right)\, ,
\label{EQ:alphasxyTMF}
\eeq

\noindent and $\mathcal{D}_{b,\uparrow,\downarrow}$ with their energy derivative are given by Eqs.(\ref{EQ:Dbdefinition}) and (\ref{EQ:Dbderivative}). We have therefore to try to get some insight into the energy dependence of $\mathcal{D}_{b,\uparrow}$ and  $\mathcal{D}_{b,\downarrow}$. Our driving idea is that the factors leading to the band contribution $\mathcal{D}_{b}$ are the same factors leading to the atomic one $\mathcal{D}_{a}$. For the atomic ones we can extend the third Hund's rule to the energy bands to grasp its behavior. The third Hund's rule for atomic states is related to the spin-orbit energy which is proportional to the product between the spin and the orbital angular momentum. With electronic orbitals that are partially unquenched and still possess an orbital angular momentum we will find Bloch states with orbital momentum with opposite sign with respect to the spin at lower energy with respect to the ones with same sign \cite{Berger-1964}. This criterion resembles the third Hund's rule for the filling of atomic orbitals but is here applied to the Bloch states. Finally, following our driving idea, we further assume that the band contribution $\mathcal{D}_{b}$ is just proportional to the atomic one $\mathcal{D}_{a}$. A sketch is shown in Fig. \ref{FIG:Bands1c} bottom. A justification of this assumption is given in Section \ref{SECTION:Atomic and band orbital momentum} on the basis of a simplified two orbitals tight binding calculation of the band structure.

\subsection{Exchange splitting}
\label{SECT:Exchange splitting}

In ferromagnets we have the formation of an atomic magnetic moment and the establishment of a long range magnetic order. The atomic moment corresponds to the presence of electrons with same spin on the same atom which is energetically advantageous in terms of a reduced shielding of the nuclear potential (the first Hund's rule, also called intra-atomic exchange). The long range order depends on the inter-atomic Heisenberg interaction between electrons. Two electrons with the same spin must be characterized by a spatially antisymmetric wavefunction, the so called antibonding states \cite{OHandley-2000}. Then, in a metallic ferromagnet, an exchange energy splitting occurs between majority and minority antibonding states and the energy gain corresponds to a reduced electrostatic repulsion between electrons. Therefore, in a simplified rigid band model at zero temperature, the filling starts from the bonding states in which majority and minority bands are filled in parallel. Then the filling proceed with the majority antibonding states giving rise to the first half of the Staler-Pauling curve in which the zero temperature magnetization is proportional to the number of $d$ electrons (the so called weak ferromagnets, like Fe). Then, when the majority $d$ band is full, the extra $d$ electrons are accommodated in the minority band and the zero temperature magnetization decreases accordingly (the so called strong ferromagnets like Co and Ni) \cite{Stearns-1986}. A sketch of the band structure is shown in Fig.\ref{FIG:Bands3dbis}. When temperature is not zero the spontaneous magnetization is reduced and the exchange splitting is decreased accordingly. However for Fe, Co and Ni the Curie temperature is large enough to allow us to employ the zero temperature band structure to interpret the transport coefficients.

\subsection{Transport coefficients of $3d$ ferromagnets}

With the assumptions made up to now one has all the elements to grasp the behavior of the transport coefficients of Eqs.(\ref{EQ:sigmaeTMF}), (\ref{EQ:epsiloneTMF}), (\ref{EQ:sigmasxyTMF}) and (\ref{EQ:alphasxyTMF}) as a function of the number of electrons in a rigid band model. Fig.\ref{FIG:Bands3dbis} shows the particular sequence of signs and amplitudes obtained for the 3$d$ ferromagnets as a function of the electron filling. It is worthwhile to compare the sequence of Fig.\ref{FIG:Bands3dbis} with the experimental values \cite{Blatt-2012, Hurd-2012,Chuang-2017}. The transport coefficients of 3$d$ bulk ferromagnets Fe, Co and Ni around room temperature are reported in Table \ref{TABLE:Spontaneous}. From experimental values we can compute the three terms of Eq.(\ref{EQ:Nerns_eq}). Of particular interest are the coefficients changing sign as the electron filling increases, i.e. $\varepsilon_e$ (Eq.(\ref{EQ:epsiloneTMF})), $\sigma_{s,xy}$ (Eq.(\ref{EQ:sigmasxyTMF})) and $-\alpha_{s,xy}$ (Eq.(\ref{EQ:alphasxyTMF})). The change of sign of the thermopower is due, in the rigid band model with exchange spitting, to the fact that the main contribution is given by the minority band for weak ferromagnets (Fe) and by the majority band for strong ferromagnets (Co and Ni). The other two coefficients are instead proportional to $\mathcal{D}_b$ and $\partial_{\epsilon}\mathcal{D}_b$ of the band being filled (majority for Fe and minority for Co and Ni). We see that the signs obtained in Table \ref{TABLE:Spontaneous} are in agreement with the scheme of Fig.\ref{FIG:Bands3dbis}. For Fe $\mathcal{D}_b$ is negative but still decreasing, for Co negative and increasing and for Ni positive and increasing. This explain the large values obtained for $\sigma_e\varepsilon_{sN}$ for Fe (negative) and Co (positive). In both cases the two contributions to the Nernst effect have the same sign. Having established a connection between the signs of the contribution and the resulting Nernst effect it is worthwhile to try a possible maximization of the effect, at least with the series of the 3$d$ ferromagnets. To this aim in the next section we put together the basic ingredients to obtain a simple predictive model.

\section{Model with two $d$ orbitals}
\label{SECTION:Model with two d orbitals}

To build a minimal model of 3$d$ transition metal ferromagnet able to account for the transport properties as described in the previous section, we take only two electronic orbital per spins with partially unquenched orbital momentum. This simplifying assumption is justified in transition metals with $d$ electrons because the crystal field energy term $\Delta \epsilon_{cf}$ produces a splitting between $d$ orbitals with different symmetry: the $e_g$-type ($d_{3z^2-r^2}$ and $d_{x^2-y^2}$) and $t_{2g}$-type ($d_{xy}$, $d_{yz}$, $d_{zx}$). The hybridization between the three $t_{2g}$ orbitals can still produce partial unquenching then preserving an orbital angular momentum ($m_l = \pm1$) \cite{Brooks-1940}. If we limit to take a specific direction for the spin, say the $z$ axis, we have only two relevant quenched orbitals, $d_{yz}$ and $d_{zx}$ or the corresponding two unquenched orbitals, $d_{+1}$ and $d_{-1}$. We therefore limit to take only two unquenched orbitals and apply the spin-orbit splitting, $\Delta \epsilon_{soc}$, to them. In this section it is appropriate to use the density of states per atom, then we replace $\mathcal{D} \rightarrow v_r \mathcal{D}$, where $v_r$ is the volume of the atomic unit cell. 

\subsection{Atomic and band orbital momentum}
\label{SECTION:Atomic and band orbital momentum}

To get a justification of our working hypothesis on the relation between the energy dependence of the atomic contribution, $\mathcal{D}_{a}$, and the band contribution, $\mathcal{D}_{b}$ to the orbital angular momentum discussed in Section \ref{SUBSECT:The third Hund's rule for Bloch states}, we show their behavior in simplified tight binging model with two bands. 

\begin{figure}[htb]
\centering
\includegraphics[width=8cm]{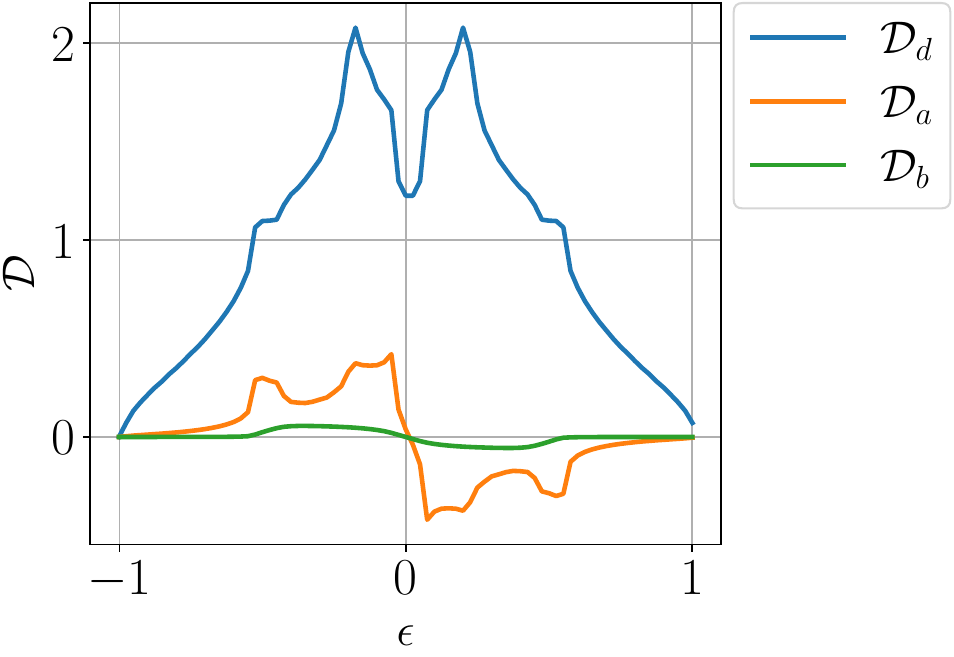}
\caption{Results of a tight binding model with the two orbitals, $d_{zx}$ and $d_{yz}$, on the bcc lattice, both with $m_s = -1/2$ (majority electrons). Density of $d$ states $\mathcal{D}_d$ (Eq.(\ref{EQ:Int_DOS_N})), density of atomic orbital angular momentum (in unit $\hslash$) $\mathcal{D}_a$ and density $\mathcal{D}_b$ of band orbital angular momentum (in unit $\hslash$) computed by Eq.(\ref{EQ:Dbdefinitionwithvolume}). The parameters of the  tight binding model are given in the main text.} \label{FIG:TBFIG}
\end{figure}

We consider a tight binding model with the $d_{zx}$ and $d_{yz}$ orbitals on a bcc lattice both with $m_s = -1/2$ (majority electrons) \cite{Kontani-2007, Kontani-2013, Dowinton-2022}. To account for the spin orbit interaction we employ the unquenched basis $d_{+1}$ and $d_{-1}$. With just two bands the tight binding Hamiltonian is written in the standard form 

\beq
\mathcal{H} = S \mathbf{1} + \mathbf{P}\cdot\boldsymbol{\sigma}
\eeq

\noindent where $\mathbf{1}$ is the unit matrix and $\boldsymbol{\sigma}$ are Pauli matrices. The matrix elements: $S$ and the components of the vector $\mathbf{P} = X \hat{\mathbf{x}}+Y \hat{\mathbf{y}}+Z \hat{\mathbf{z}}$ are all functions of the wavevector $\mathbf{k}$. After diagonalization we find the energy of the two bands as $\epsilon_{\pm} = S \mp P$ and their Berry curvature (in the case the spin is along $z$ and $Z$ is a constant) as 

\beq
\Omega_{\pm} = \pm \frac{1}{2} \frac{Z}{P^3} \left( \partial_{k_x} X \partial_{k_y} Y - \partial_{k_y} X \partial_{k_x} Y\right)
\eeq

\noindent as shown by Eq.(14) of Ref.\cite{Graf-2021}. The matrix elements can be computed from the hopping integrals by the help of the Slater-Koster expressions \cite{Slater-1954}, using the relation between quenched and unquenched orbitals $d_{yz} = i(d_{+1}+ d_{-1})/\sqrt{2}$ and $d_{zx}=(-d_{+1}+d_{-1})/\sqrt{2}$ and by considering both first neighbor and second neighbor interaction. For the bcc lattice with $k_x^{\prime} = (k_x+k_y)/2$, $k_y^{\prime} = (-k_x+k_y)/2$ and $k_y^{\prime}=k_z$ in the range $[-\pi:\pi]$ one obtains a cubic Brillouin zone. By dropping the prime we can write 

\bea
S &=&  - \gamma_{1m} \frac{1}{2} \cos\left(\frac{k_z}{2}\right) \left[ \cos\left(k_x\right) + \cos\left(k_y\right)\right]\\
X &=& - \gamma_{2} \sin(k_x)\sin(k_y)\\
Y &=& \gamma_{1p} \frac{1}{2} \cos\left(\frac{k_z}{2}\right) \left[ - \cos\left(k_x\right) + \cos\left(k_y\right)\right]
\eea

\noindent where the coefficients $\gamma_{1p}$, $\gamma_{1m}$ and $\gamma_2$ are related to the Slater hopping integrals

\bea
\gamma_{1m} &=& \frac{8}{3} \left[  \gamma_{dd\sigma,1} - \frac{2}{3} \left( \gamma_{dd\pi,1} - 2 \gamma_{dd\delta,1}\right)\right]\\
\gamma_{1p} &=& \frac{8}{3} \left[ \gamma_{dd\sigma,1} +  \frac{1}{3} \left(\gamma_{dd\pi,1} - 2 \gamma_{dd\delta,1}\right)\right]\\
\gamma_2 &=& 2 (\gamma_{dd\pi,2} + \gamma_{dd\delta,2} )
\eea

\noindent and the subscripts 1 and 2 refers to first and second neighbor interaction. The coefficient $Z$ is due to the spin orbit interaction. Appropriate normalized values are $\gamma_{1m} \simeq 0.5 $, $\gamma_{1p} \simeq 1$, $\gamma_2 \simeq 0.2$, $Z \simeq - 0.05$ \cite{Ducastelle-1970}. The density of $d$ states $\mathcal{D}_d$ is given by Eq.(\ref{EQ:Int_DOS_N}). As we are working with the unquenched orbital basis we know that the $z$ component of the atomic contribution to the orbital momentum is $l_a(d_{+1}) = 1$ and $l_a(d_{-1}) = -1$ (in unit $\hslash$). Therefore for the two bands we get $l_{a}(\pm) = \mp Z/P$. The density $\mathcal{D}_b$ of band orbital angular momentum (in unit $\hslash$) is computed by Eq.(\ref{EQ:Dbdefinitionwithvolume}). The results are shown in Fig.\ref{FIG:TBFIG}. The hybridization between the two orbitals gives rise to $\mathcal{D}_a$ and $\mathcal{D}_b$ with the same signs in the same regions of the energy spectrum. It should be noted that the presence of the spin orbit term $Z$ is essential to obtain unquenched orbitals giving rise to a non zero atomic orbital momentum, $\mathcal{D}_a$, and an additional presence of non zero $\gamma_2$ coefficient gives a non zero Berry curvature and a non zero band orbital momentum $\mathcal{D}_b$. Further developments of the two bands models will be presented in a subsequent paper \cite{DiPietro-2025}.

\subsection{Analytical model for $3d$ metals}

\begin{figure*}
    \centering
    \includegraphics[width=18cm]{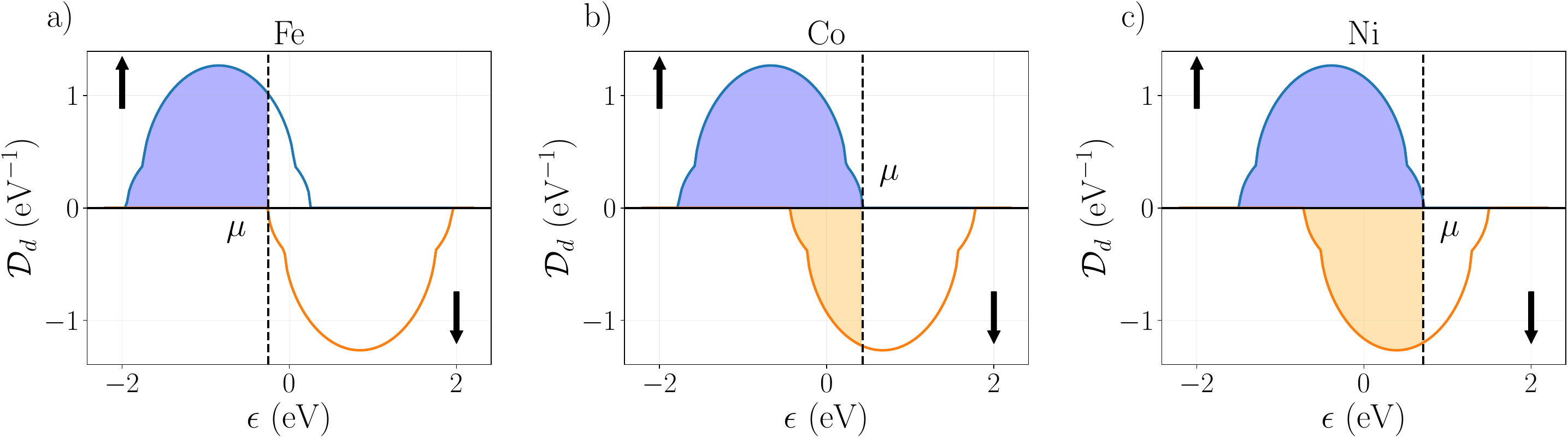}
    \caption{Density of $d$-states for the model with two orbitals of Eqs.(\ref{EQ:Dtwoorbitalsup}) and (\ref{EQ:Dtwoorbitalsdown}). The picture shows the majority (blue) and minority (orange) spin bands and the three cases of electron filling i.e. a) Fe ($\mathcal N_d = 1.7$) , b) Co ($\mathcal N_d = 2.7$) and c) Ni ($\mathcal N_d = 3.4$). Band structure parameters: $\Delta \epsilon_d = 1.0 ~ \mathrm{eV}; ~ \Delta \epsilon_{soc} = 0.1 ~ \mathrm{eV}$.}
    \label{FIG:DOS_filling}
\end{figure*}

\begin{figure}
    \centering
    \includegraphics[width=8cm]{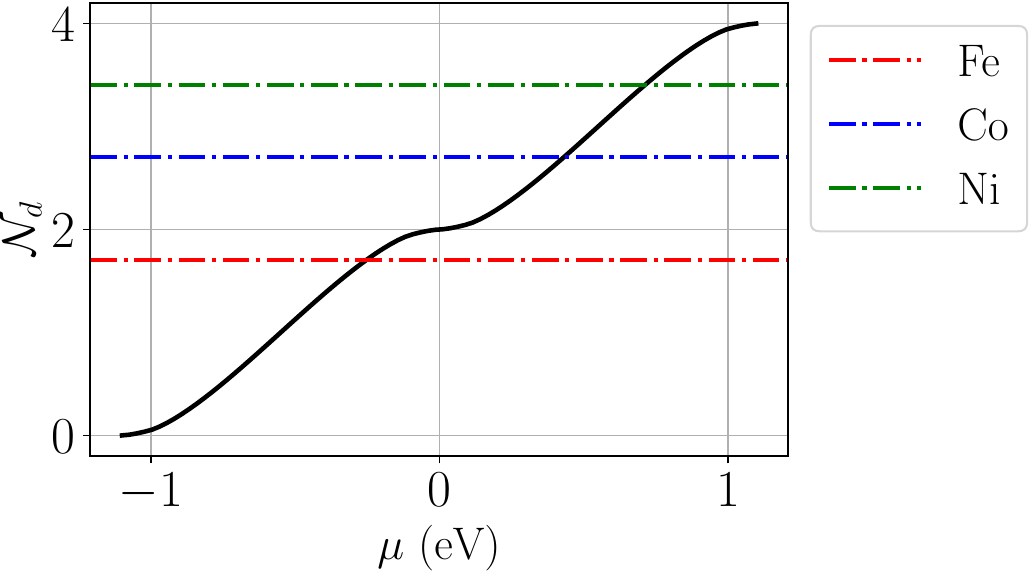}
    \caption{Electron filling $\mathcal N_d$ as a function of the Fermi level $\mu$. The red, blue and green horizontal dashed lined represent the electron filling levels of Fe ($\mathcal N_d = 1.7$), Co ($\mathcal N_d = 2.7$) and Ni ($\mathcal N_d = 3.4$) respectively.}
    \label{FIG:Filling}
\end{figure}

With the aim to derive an approximate rigid band model for $3d$ ferromagnets, we generalize the previous result valid for a specific lattice model, and assume reasonable shapes for the densities $\mathcal{D}_{d}$, $\mathcal{D}_{a}$ and $\mathcal{D}_{b}$ in order to compute the transport properties. We assume a density of states $\mathcal{D}(\epsilon)$ for each of the two $d$ orbitals and take $\mathcal{D}_{d}$ as the superposition of the spin orbit energy split bands: 

\beq
\mathcal{D}_{d}(\epsilon) = \mathcal{D}_{+}(\epsilon)+\mathcal{D}_{-}(\epsilon)
\eeq

\noindent where $\mathcal{D}_{\pm}(\epsilon) = \mathcal{D}(\epsilon \pm \Delta \epsilon_{soc})$ refers to the $d_{+1}$ and $d_{-1}$ orbital states and the distribution $\mathcal{D}(\epsilon)$ is normalized to one $\int_{- \infty}^{\infty} \mathcal{D} \, d \epsilon = 1$. $\mathcal{D}(\epsilon)$ is assumed to be a semi-circular distribution $\mathcal{D}(\epsilon) = (2/\pi) [ 1 - (\epsilon/\Delta \epsilon_d)^2]^{1/2}/\Delta \epsilon_d$. This distribution, behaving $ \sim \sqrt{\epsilon}$ at the band edges, is widely used in dynamical mean field theory treatments \cite{Georges-1996, Economou-2006}. The parameter $\Delta\epsilon_0 = \Delta\epsilon_d+\Delta\epsilon_{soc}$, i.e. the half of the width of the band $\mathcal{D}_{d}$, is the energy scale of the problem. In the ferromagnetic state the spin polarized bands are split by the exchange energy $\Delta \epsilon_{ex}$: 

\bea
\mathcal{D}_{d,\uparrow} &=& \mathcal{D}_{d}\left(\epsilon + \Delta \epsilon_{ex} \right) \, , 
\label{EQ:Dtwoorbitalsup}
\\
\mathcal{D}_{d,\downarrow} &=& \mathcal{D}_{d}\left(\epsilon - \Delta \epsilon_{ex} \right) \, .
\label{EQ:Dtwoorbitalsdown}
\eea

\noindent In agreement with the discussion of Section \ref{SECT:Exchange splitting} we assume that at zero temeperature the exchange splitting parameter $\Delta \epsilon_{ex}$ depends on the electron filling of the $d$ band as 

\beq
\Delta \epsilon_{ex} = \Delta \epsilon_{0} - |\mu| \, .
\label{EQ:Deltae_ex}
\eeq

\noindent With the above mentioned assumptions the number of $d$ electrons per atom is $\mathcal{N}_d = \int_{- \infty}^{\mu} ( \mathcal{D}_{d,\uparrow} + \mathcal{D}_{d,\downarrow} ) ~ d \epsilon$ and ranges from $0$ to $4$, i.e.: from 0 to 2 for weak ferromagnets (like Fe) and from 2 to 4 for strong ferromagnets (like Co and Ni). In such a simplified rigid band model of transition metal magnetism we allow 0.3 holes for Fe in the majority band, 0.7 electrons for Co and 0.6 holes for Ni in the minority band \cite{OHandley-2000}. Fig.\ref{FIG:DOS_filling} shows the corresponding density of $d$-states for Fe, Co and Ni and Fig.\ref{FIG:Filling} shows $\mathcal{N}_d$ as a function of the Fermi level $\mu$. The final density of states will also include the contribution of $s,p$ electrons as $\mathcal{D}_{N} = \mathcal{D}_{d}+\mathcal{D}_{s}$ where $\mathcal{D}_{s}$, associated to a much wider energy band, is taken as a constant. 

For the density of atomic orbital momentum (in unit $\hslash$) we also take the superposition of spin orbit energy split atomic orbital distributions, but we have to account for the different signs, so that 

\bea
\mathcal{D}_{a,\uparrow } &=&  \mathcal{D}_{l, +,\uparrow}  - \mathcal{D}_{l,-,\uparrow} \, , 
\label{EQ:Daup} \\ 
\mathcal{D}_{a,\downarrow }  &=& - \mathcal{D}_{l, +,\downarrow} + \mathcal{D}_{l,-,\downarrow} \,. 
\label{EQ:Dadown} 
\eea

\noindent In Fig.\ref{FIG:TBFIG} we observe that the densities of atomic and band orbital momentum (in unit $\hslash$), $\mathcal{D}_{a}$ and $\mathcal{D}_{b}$ have a different functional behavior at the band edges with respect to $\mathcal{D}_{d}$ therefore we assume that the individual orbital distribution $\mathcal{D}_{l}(\epsilon)$ is smoothed at the band edges. We assume $\mathcal{D}_{l}(\epsilon) = s(\epsilon)\mathcal{D}(\epsilon)$ where $s(\epsilon)$ is the smoothing function taken as $s(\epsilon) = (1-\cosh(\epsilon/\eta)/\cosh(\Delta \epsilon_d/\eta))^2$. Finally the density of itinerant angular momentum $\mathcal{D}_{b}$ is taken to be proportional to $\mathcal{D}_{a}$ i.e. 

\beq
\mathcal{D}_{b} = b \mathcal{D}_{a}
\label{EQ:Db_tot}
\eeq

\noindent where $b$ is a dimensionless coefficient $b < 1$. As we have seen in the previous paragraph both $\mathcal{D}_{a}$ and $\mathcal{D}_{b}$ can be determined by means of the hopping integrals in a tight binding approach \cite{DiPietro-2025}. However, in the present context, $b$ is taken an adjustable parameter.  

\subsection{Reduced transport coefficients}
\label{SUBSEC:Normalization}

To derive the $d$-band parameters $\Delta \epsilon_d$, $\Delta \epsilon_{soc}$ and $\mathcal{D}_s$, $\eta$ and $b$ we divide the transport coefficients of Eqs.(\ref{EQ:sigmaeTMF}), (\ref{EQ:epsiloneTMF}), (\ref{EQ:sigmasxyTMF}) and (\ref{EQ:alphasxyTMF}) by the known constants:

 \begin{align}
     \hat{\sigma}_{e} &= \sigma_{e}/(G_0 c/\Delta\epsilon_d), \\
     \hat{\varepsilon}_{e} &= \varepsilon_{e}/(\varepsilon_{e,0} t \Delta\epsilon_d), \\
     \hat{\sigma}_{s,xy} &= \sigma_{s,xy}/(G_0 d \Delta\epsilon_d), \\
     \hat{\alpha}_{s,xy} &= \alpha_{s,xy}/(\varepsilon_{e,0} t G_0 d \Delta\epsilon_d^2),
 \end{align}

\noindent where $G_0 = (e^2/h) \simeq 38.7\cdot 10^{-6} \, \Omega^{-1}$,  $\varepsilon_{e,0} = (k_B/e) \simeq 86 \cdot 10^{-6}$ V K$^{-1}$, $c = ((1/3)S_Fv_F v_r /f_c)\Delta\epsilon_d$ is the collision parameter (unit m$^{-1}$), $ d = (2\pi \hslash^2 /m_e v_r)/\Delta \epsilon_d$ is the displacement parameter (unit m$^{-1}$) (for transition metals we use $v_r \simeq 1.1 \cdot 10^{-29}$ m$^{3}$ and get $d\Delta \epsilon_d \simeq 4.34 \cdot 10^{10}$ eV m$^{-1}$), $t = (\pi^2/3) k_BT/\Delta \epsilon_d$ is the normalized temperature ($\Delta \epsilon_d t=0.082$ eV at $T = 296$ K) and write the reduced coefficent as

\beq
\hat{\sigma}_{e} = \frac{1}{\mathcal{D}_{N,\uparrow}} +\frac{1}{\mathcal{D}_{N,\downarrow}} \, , 
\label{EQ:hat_sigma_e}
\eeq

\beq
\hat{\varepsilon}_{e} = \frac{\mathcal{D}_{N,\downarrow}   \partial_{\epsilon} \ln\mathcal{D}_{N,\uparrow}
 + \mathcal{D}_{N,\uparrow}   \partial_{\epsilon} \ln \mathcal{D}_{N,\downarrow}
 }{\mathcal{D}_{N,\uparrow} + \mathcal{D}_{N,\downarrow}} \, ,
 \label{EQ:hat_epsilon}
\eeq

\beq
\hat{\sigma}_{s,xy} = \mathcal{D}_{b,\uparrow}+\mathcal{D}_{b,\downarrow} \, ,
\label{EQ:hat_sigma_hall}
\eeq

\noindent and 

\beq
- \hat{\alpha}_{s,xy} = \frac{\partial \mathcal{D}_{b,\uparrow}}{\partial \epsilon}+ \frac{\partial \mathcal{D}_{b,\downarrow}}{\partial \epsilon} \, .
\label{EQ:hat_alpha}
\eeq

\noindent To deal with the energy derivatives at the kink points of the DOS (observe the band edges in Fig.\ref{FIG:DOS_filling}), we recall how our model assumes the system to be at room temperature, implicitly setting a minimal energy resolution of $k_B T$ for the DOS. We therefore define the energy derivatives such as those found in e.g. Eq.\eqref{EQ:hat_epsilon} as the finite difference

\beq
\partial_\epsilon \mathcal{D}_N = \frac{\mathcal{D}_N(\epsilon +k_BT) - \mathcal{D}_N(\epsilon  -k_BT)}{2 k_B T} \, .
\eeq

\noindent The corresponding reduced experimental values of Table \ref{TABLE:Spontaneous} are reported in Table \ref{TABLE:Spontaneousnormalized}. 

\subsection{Results}

\begin{figure}
    \centering
    \includegraphics[width=7.5cm]{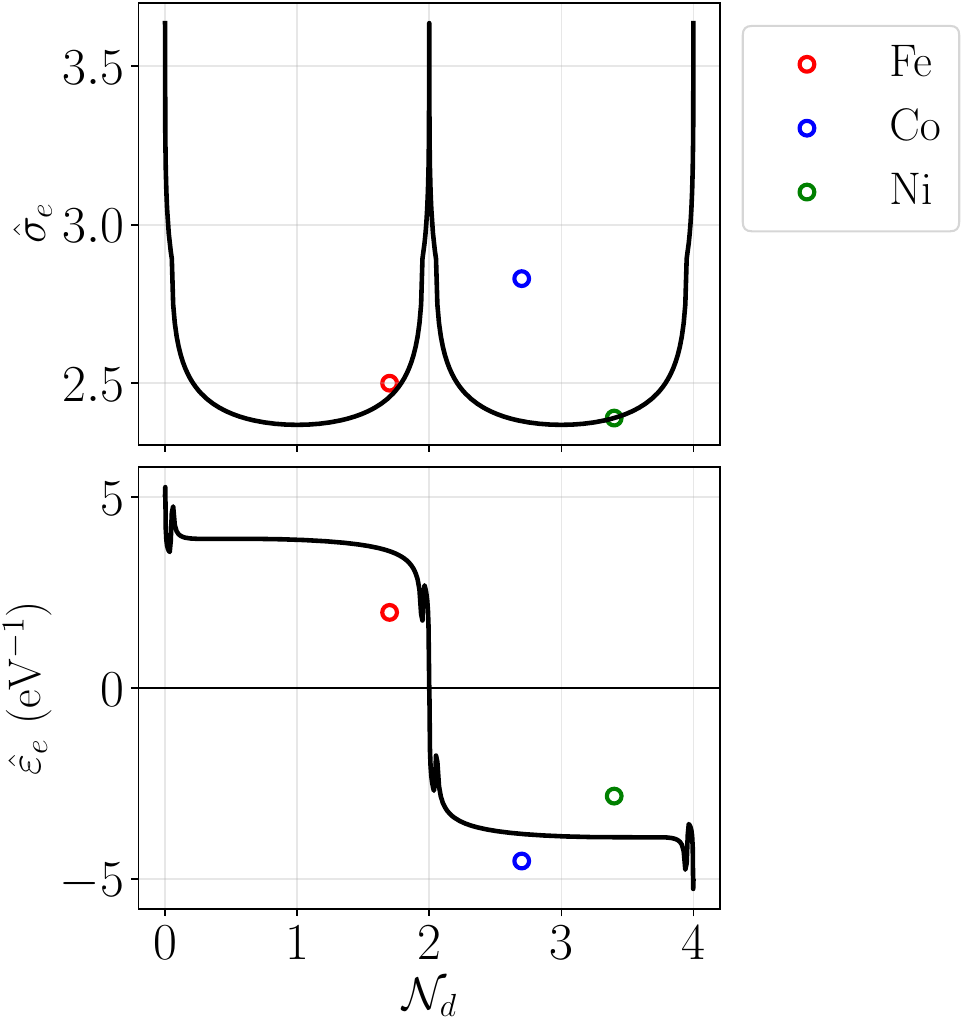}
    \caption{Model predictions of the longitudinal transport coefficients. Lines: theory: a) longitudinal electrical conductivity $\hat \sigma_e$ (Eq.\eqref{EQ:hat_sigma_e}) and b) thermopower coefficient $ \hat \varepsilon_e$ (Eq.\eqref{EQ:hat_epsilon}). Symbols: experimental values from Table \ref{TABLE:Spontaneousnormalized}.}
    \label{fig:Conductivity_Seebeck}
\end{figure}

A good agreement between the model of Eqs.(\ref{EQ:hat_sigma_e}-\ref{EQ:hat_alpha}) and the experimental values of Table \ref{TABLE:Spontaneousnormalized} is obtained by setting $\Delta \epsilon_{d} = 1.0 \,\mathrm{eV}$, $\Delta\epsilon_{soc} = 0.1 \, \mathrm{eV}$, $\mathcal{D}_s = 0.55 ~\mathrm{eV}^{-1}$ and the collision parameter $ c/\Delta \epsilon_d = 2.8 \cdot 10^{10} \,$ eV$^{-1}$m$^{-1}$. 

The conductivity $\hat{\sigma}_e$ of Eq.(\ref{EQ:hat_sigma_e}) is shown in Fig \ref{fig:Conductivity_Seebeck}a. We observe that at the band edges ($0$ and $2$ electron filling each of the two sub-bands) the conductivity is completely supported by the $s$-band and reaches the value of $3.6$, which corresponds to $2/\mathcal{D}_s$. As soon as the $d$-band start to fill up, the conductivity drops because of the increased scattering due to the $d$-states. We can see how all three levels of filling relative to Fe, Co and Ni correspond to similar values for the conductivity, a fact that is also reflected in Table \ref{TABLE:Spontaneousnormalized}, reporting experimental data. 

The thermopower $\hat{\varepsilon}_e$ of Eq.(\ref{EQ:hat_epsilon}) is reported in Fig \ref{fig:Conductivity_Seebeck}b. We observe a characteristic change of sign as we cross the filling of the majority band and proceed to the filling of the minority band. The switch is related to the derivative sign at the edge of the minority (large positive derivative) and majority electrons (large negative derivative) which ends up being the dominating contribution for this effect in our simplified model (see Fig.\ref{FIG:DOS_filling}). A comparison with the experimental data of Table \ref{TABLE:Spontaneousnormalized} reveals agreement with the signs of the coefficients.

\begin{figure}
    \centering
    \includegraphics[width=8.5cm]{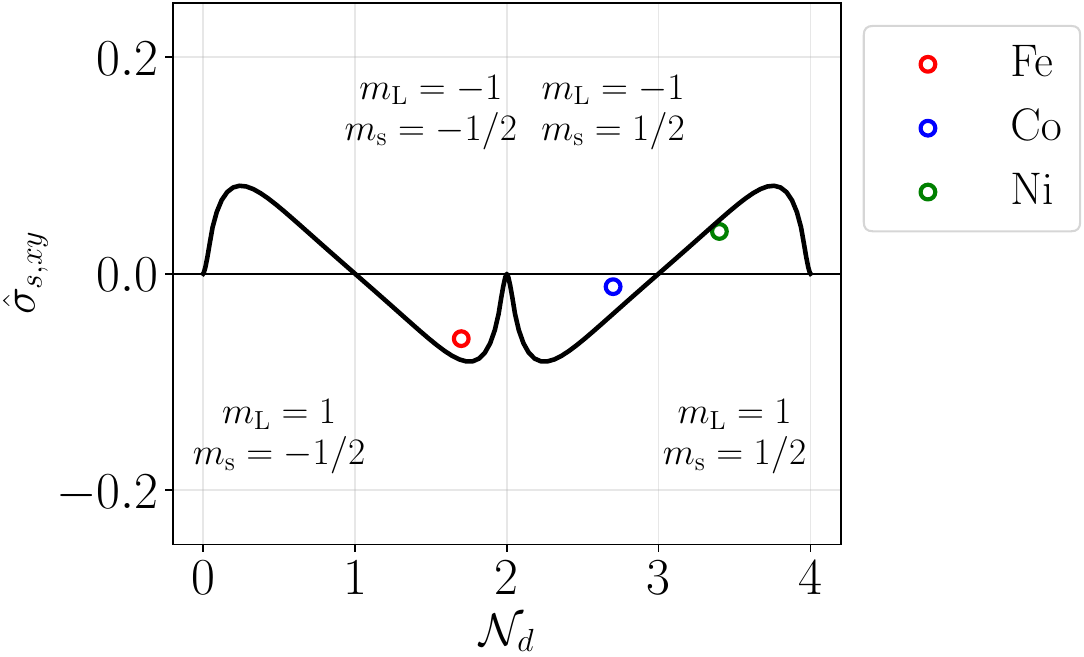}
    \caption{Model predictions of the spontaneous Hall conductivity $\hat{\sigma}_{s,xy}$. Lines: theory of Eq.(\ref{EQ:hat_sigma_hall}) as a function of $d$ electron filling. Symbols: experimental values on Fe, Co and Ni from Table \ref{TABLE:Spontaneousnormalized}.}
    \label{fig:sigma_s_xy}
\end{figure}

The spontaneous Hall conductivity $\hat \sigma_{s, xy}$ of Eq.\eqref{EQ:hat_sigma_hall} is shown in Fig.\ref{fig:sigma_s_xy}. In the lower part of the majority band ($0<\mathcal N_d < 1$), the band orbital moment is aligned to the majority spin ($d_{+1,\uparrow}$ states) and therefore has positive values. As we cross $\mathcal N_d = 1$ the band orbital moment changes sign. As soon as we start filling the minority band, the first states that start to be occupied are in the $d_{-1,\downarrow}$ state, because SOC ($\Delta \epsilon_{soc}$) pushes states aligned with the spin lower in energy. Once the first half of the minority band is filled ($3 \le \mathcal N_d < 4$), the band orbital moment density changes sign again. Since the $\eta$ and $b$ parameter introduced in Eq.\eqref{EQ:Db_tot} are purely phenomenological in the present model, we set them to $\eta = 0.35$ and $b=0.5$ in order to approximately align the theoretical values obtained with Eq.\eqref{EQ:hat_sigma_hall} with the dimensionless experimental values $\hat \sigma_{s, xy}$ of Table \ref{TABLE:Spontaneousnormalized}. We now proceed and plot all the remaining thermomagnetic coefficients without any additional adjustable parameter.

\begin{figure*}
    \centering
    \includegraphics[width=17 cm]{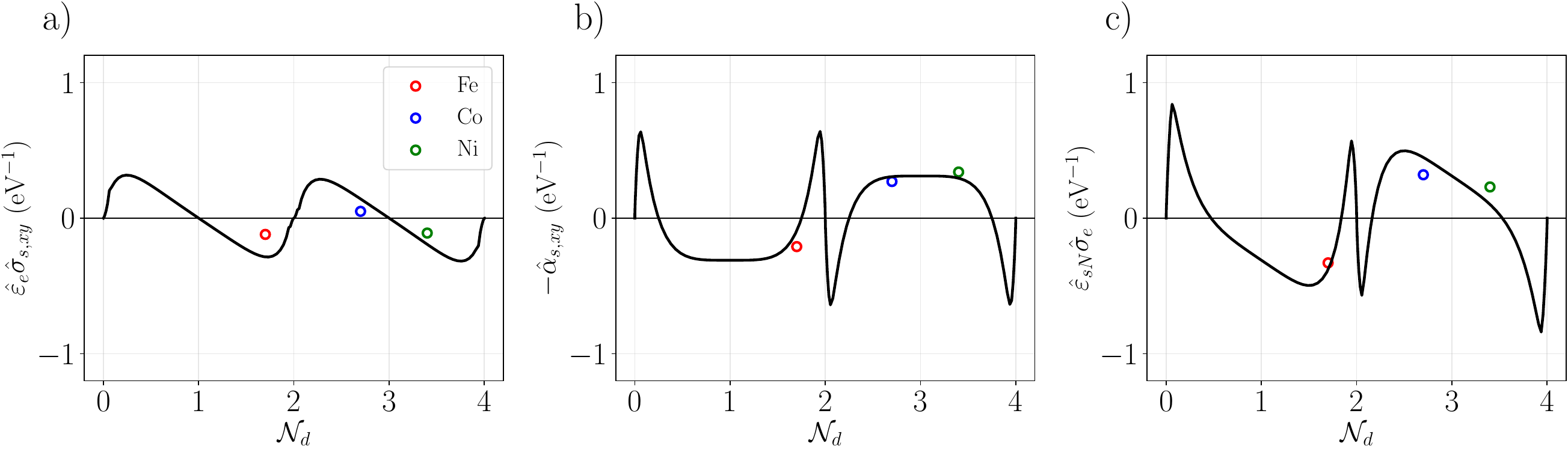}
    \caption{Model predictions of the transverse transport coefficients. Lines: theory from Eqs.(\ref{EQ:hat_epsilon}), (\ref{EQ:hat_sigma_hall}) and (\ref{EQ:hat_alpha}). a) Combined effect of $ \hat{\varepsilon}_e \hat{\sigma}_{s,xy}$, b) spontaneous Nernst conductivity $ -\hat{\alpha}_{s,xy}$ and c) spontaneous Nernst coefficient multiplied by the electron conductivity: $\hat{\varepsilon}_e \hat{\sigma}_{s,xy} - \hat{\alpha}_{s,xy}$. Symbols: experimental values on Fe, Co and Ni as reported in Table \ref{TABLE:Spontaneousnormalized}.}
    \label{fig:Nernst_final}
\end{figure*}

Fig. \ref{fig:Nernst_final} show: a) the product $\hat \varepsilon_{e} \hat \sigma_{s,xy}$, b) $ - \hat \alpha_{s,xy}$ and c) the product $\hat \varepsilon_{sN} \hat \sigma_e$  which is given by the sum of a) and b). All this plots are odd functions with respect to the half filling point $\mathcal N_d = 2$. Observing Fig.\ref{fig:Nernst_final}a we notice how the multiplication of the thermopower and the spontaneous Hall conductivity $ \hat \sigma_{s,xy} \hat \varepsilon_e$ provides a contribution which is changing sign during the filling of each spin band whereas $-\hat{\alpha}_{s,xy}$ (Fig. \ref{fig:Nernst_final}b) given by the energy derivative of the orbital moment density, displays a plateau in the middle of the majority and minority band filling (i.e. in the regions $0 \lesssim \mathcal N_d \lesssim 2 $  and $2 \lesssim \mathcal N_d \lesssim 4$). This is related to the slope of the orbital moment density in those regions (see Fig.\ref{fig:sigma_s_xy}), which is almost constant in our approximation. The sign of the spontaneous Nernst coefficient contained in the product $ \hat \varepsilon_{sN} \hat \sigma_e $ (Fig. \ref{fig:Nernst_final}c) is largely determined by the sign of the plateau of $- \hat \alpha_{s,xy}$, which is negative for weak ferromagnets and positive for strong ferromagnets. Despite the strong approximations of this model we observe how the three components of Fig.\ref{fig:Nernst_final}, $\hat \varepsilon_{sN} \hat \sigma_e = \hat \varepsilon_{e} \hat \sigma_{s,xy} - \hat \alpha_{s,xy}$ are well superposed to the experimental ones (see Table \ref{TABLE:Spontaneousnormalized}). We therefore expect that with the obtained model, it is possible to make some statements regarding maximization strategies of the spontaneous Nernst coefficient. 

\begin{figure}[htb]
\centering
\includegraphics[width=7cm]{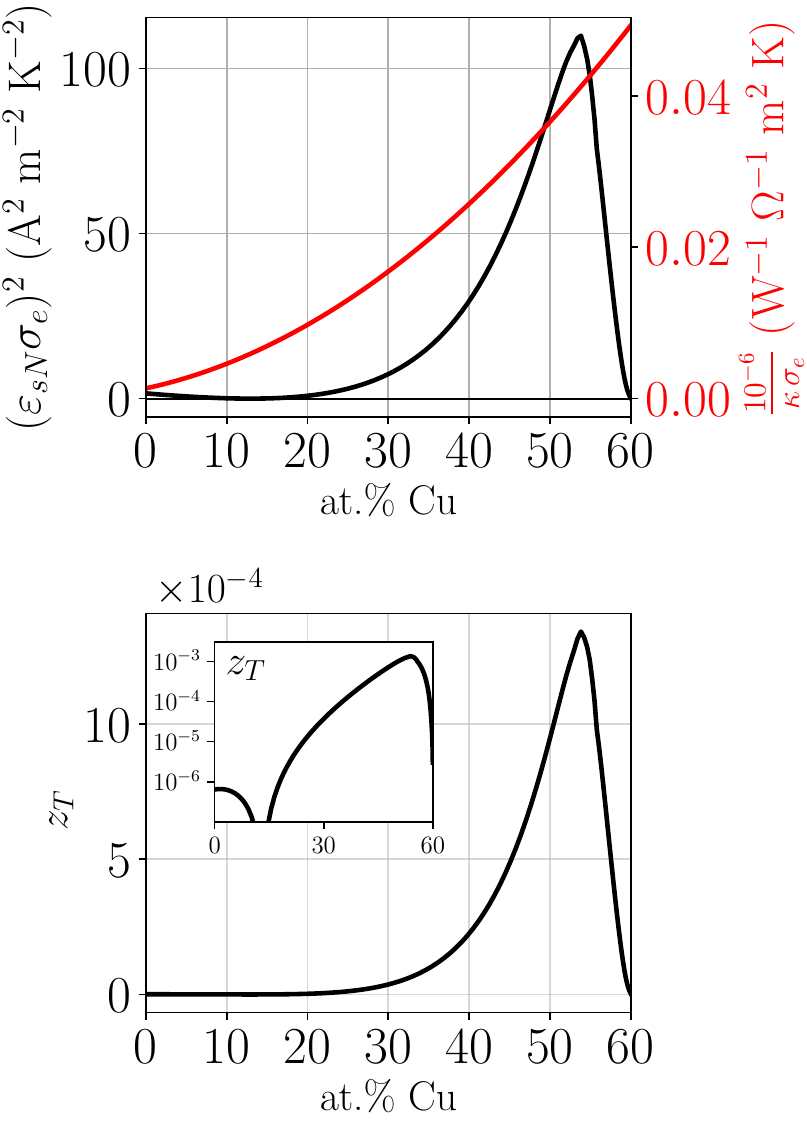}
\caption{Model prediction of the behavior of Ni$_{1-x}$Cu$_x$ alloys. Top: product $(\varepsilon_{sN} \sigma_e)^2$ taken from $\hat \varepsilon_{sN} \hat \sigma_e$ of Fig.\ref{fig:Nernst_final}c using $x = N_d - 3.4$ and inverse of the product of the thermal conductivity $\kappa$ and the electric conductivity as $\sigma_{e}$ as a function of composition $x$. Bottom. Predicted values of $z_T$ versus $x$ from Eq.(\ref{EQ:zT}) by using the values of the top plot. The inset shows the same result in logarithmic scale.} \label{FIG:zTCu}
\end{figure}

As visible in Fig.\ref{fig:Nernst_final}c, a viable maximization strategy could be to push the Fermi level as close as possible to the complete filling ($\mathcal N_d \approx 4$) or emptying ($\mathcal N_d \approx 0$). In these regions, the contributions from the spontaneous Hall and Nernst conductivities have the same signs (observe Figs. \ref{fig:Nernst_final}a and \ref{fig:Nernst_final}b) and act constructively. For example, at $\mathcal N_d \simeq 3.93$, we find a maximum of $\hat \varepsilon_{sN} \hat \sigma_e \simeq -0.85$ eV$^{-1}$. Among the many transition metal alloys in the $3d$-series close to the complete filling of the minority band, we find the Ni$_{1-x}$Cu$_x$ alloys. In this regard, it is worth to mention that the Ni$_{70}$Cu$_{30}$ alloy has recently been shown to show interesting spin orbit effect as for example a large spin Nernst effect \cite{Li-2025}. Taking the result of $\hat \varepsilon_{sN} \hat \sigma_e$ from Fig.\ref{fig:Nernst_final}c, using $x = N_d-3.4$ and taking the composition dependence of the thermal conductivity and the electric conductivity as $\sigma_{e}(x) = \sigma_{e}(0)(1+\alpha_{\sigma}x)^{-1}$ and $\kappa(x) = \kappa(0)(1+\alpha_{\kappa}x)^{-1}$ with $\sigma_{e}(0)$ and $\kappa(0)$ taken from Table \ref{TABLE:Spontaneous} (Ni) and $\alpha_{\sigma}=10$ and $\alpha_{\kappa}=7$ we get the plot of Fig.\ref{FIG:zTCu} top. The predicted values of $z_T$ versus $x$ are shown in Fig.\ref{FIG:zTCu} bottom. The curve shows a drastic improvement with respect to pure nickel ($z_T = 3.2 \cdot 10^{-6}$) when $x$ is increased, reaching a maximum of around $z_T \simeq 1.32 \cdot 10^{-3}$ at $x \simeq 0.53$. It is worth to mention that this improvement is obtained by acting on both sources of the spontaneous Nernst effect: i) an improved scattering, due to the disordered alloy structure of the metal and ii) an optimized band structure contribution. However close to the complete filling of the minority band not only the spontaneous, zero temperature, magnetic moment per atom $M(0,x)$, decreases (approximately as $M(0,x) = 0.6-x$ Bohr magnetons per atom) but also the Curie temperature decreases (approximately as $T_c(x) = T_c(0) (1-x/0.6)$). Therefore the Curie temperature reaches room temperature already at compositions $x \simeq 0.3$. A more precise prediction should be therefore be able to take into account the decrease of the exchange splitting $\Delta \epsilon_{ex}$ as the temperature gets closer to the Curie point, then extending the simple zero temperature exchange splitting of Eq.(\ref{EQ:Deltae_ex}) by considering for example a computation of the splitting as $\Delta \epsilon_{ex} = m(\Delta \epsilon_{0} - |\mu|)$ where $m$ is the reduced magnetization, a parameter equal to 1 for zero temperature and 0 at $T=T_c$. This extension will be the focus of future developments of the presented model.

\begin{table*}
\begin{center}
\begin{tabular}{|c||c|c|c||c|c|c||c||}
	\hline
& $\hat{\sigma}_{e}$ & $\hat{\varepsilon}_{sN}$ & $\hat{\sigma}_{e}\hat{\varepsilon}_{sN}$& $\hat{\sigma}_{s,xy}$ & $\hat{\varepsilon}_{e}$  & $\hat{\sigma}_{s,xy}\hat{\varepsilon}_{e}$ & $-\hat{\alpha}
_{s,xy}$ \\
&  & [eV$^{-1}$] & [eV$^{-1}$] & & [eV$^{-1}$] & [eV$^{-1}$] & [eV$^{-1}$] \\
	\hline
Fe  &4.45 & $-$0.074 &$-$0.33& $-$0.060   & 2.0     &  $-$0.12 & $-$0.21 \\
Co  & 5.03 & 0.063 &0.32           & $-$0.012 & $-$4.5  &      0.05    & 0.27\\
Ni   & 4.25  & 0.054  & 0.23        &      0.039 & $-$2.9  &  $-$0.11   & 0.34\\
	\hline
\end{tabular}
\end{center}
\caption{Reduced experimental values of the thermomagnetic coefficients (from Table \ref{TABLE:Spontaneous}) as described in Sect.\ref{SUBSEC:Normalization}.}
\label{TABLE:Spontaneousnormalized}
\end{table*}

\section{Conclusions}

In this paper we use the Botzmann formalism, combined with the Berry curvature theory of Bloch electrons in solids to set up theoretical framework able to predict the energy dependence of all thermoelectric coefficients determining the strength of the spontaneous Nernst coefficient $\varepsilon_{sN}$. As a first qualitative consequence of this theory, we observe how the spontaneous Nernst coefficient is inversely proportional to the electron scattering time $\tau_c$, in contrast to the ordinary effect. Making use of Mott’s two currents model \cite{Mott-1945, Coey-2010}, we write the energy dependence of all the thermoelectric coefficients explicitly. We show how the spontaneous Hall conductivity $\sigma_{s,xy}$ as well as the intrinsic Nernst conductivity $\alpha_{s,xy}$ are determined by the band orbital momentum density $\mathcal D_b$ and its energy derivative near the Fermi level. In particular, we explicitly show how the intrinsic Nernst conductivity $\alpha_{s,xy}$ is related to the Fermi surface integral of the Berry curvature field $\boldsymbol{\Omega}_n$. Finally, we construct a minimal model for the DOS and the density of itinerant orbital magnetic moment in $3d$ ferromagnetic metals using the semi-circular DOS approximation \cite{Georges-1996, Economou-2006} accounting for the presence of SOC and exchange splitting. This model allows to explicitly write and plot all the thermomagnetic coefficients as a function of electron filling of the $d$ bands. Despite its simplicity, the model is able to correctly predict the signs and orders of magnitudes of all the relevant thermoelectric coefficients necessary for the computation of $\varepsilon_{sN}$. With the obtained model, we are able to semi-quantitatively study regions of the DOS in which the different coefficients of Eq.(\ref{EQ:Nerns_def0}) act constructively, speculating on some possible optimization strategies for the magnitude and sign of the spontaneous Nernst coefficient $\varepsilon_{sN}$. We envision an extension of the method derived in this paper applicable to other classes of metals by replacing the rigid band model with more accurate tight binding and DFT calculations  (e.g. Ref.\cite{Stejskal-2023}) together with appropriate scattering models.

\appendix

\section{Fermi golden rule}

The Fermi golden rule expresses the fact that scattering rate depends not only on the scattering sources but also on the density of available states after the scattering event \cite{Mott-1945, Solyom-2007}. Consider the scattering of an electron initially in state $\mathbf{k}$ into a state $\mathbf{k}^{\prime}$. The scattering rate is

\beq
\frac{1}{\tau_c(\mathbf{k})} = \frac{v_r}{(2\pi)^3} \int_{v_k} P_{\mathbf{k},\mathbf{k}^{\prime}} (1-\cos\theta)d^3k^{\prime}
\eeq

\noindent where the integral extends over the Brillouin zone, $\theta$ is the angle formed by the velocities $\mathbf{v}_{\mathbf{k}}$ and $\mathbf{v}_{\mathbf{k}^{\prime}}$, the factor $(1-\cos\theta)$ describes the different weights of forward and backward scattering and $P_{\mathbf{k},\mathbf{k}^{\prime}}$ is the transition rate. The perturbation theory permits to compute $P_{\mathbf{k},\mathbf{k}^{\prime}}$ between states with the same energy as

\beq
P_{\mathbf{k},\mathbf{k}^{\prime}} = \frac{2\pi}{\hslash}|U_{\mathbf{k},\mathbf{k}^{\prime}}|^2\delta(\epsilon(\mathbf{k})-\epsilon(\mathbf{k}^{\prime}))
\eeq

\noindent where $U_{\mathbf{k},\mathbf{k}^{\prime}}= \bra{\psi_{\mathbf{k}^{\prime}}}U_s\ket{\psi_{\mathbf{k}}}$ is computed by using the scattering potential $U_s$ between the states of the unperturbed Hamiltonian. The scattering rate is then a Fermi surface integral

\beq
\frac{1}{\tau_c(\mathbf{k})} = \frac{1}{(2\pi)^2}\int_{\Sigma_F} \frac{v_r |U_{\mathbf{k},\mathbf{k}^{\prime}}|^2}{\hslash^2}  \frac{(1-\cos\theta)}{v_g(\mathbf{k}^{\prime})}  d^2k^{\prime}
\eeq

\noindent With isotropic $U_{\mathbf{k},\mathbf{k}^{\prime}} = U_{k}$, the scattering rate is found to be given by the product 

\beq
\frac{1}{\tau_c} = f_c \mathcal{D}_N
\eeq

\noindent where $\mathcal{D}_N$ is the density of states of Eq.(\ref{EQ:Int_DOS_N}) depending on the energy and $f_c$ is a scattering strength (with unit J m$^3$s$^{-1}$) 

\beq
f_c  = \frac{|U_k|^2 }{\hslash^2}hv_r
\eeq

\noindent depending on the temperature. The scattering strength $f_c$ can be computed for the main scattering processes such as: i) the scattering at defects and crystal imperfections (Ref.\cite{Solyom-2007} page 387), giving rise to the residual resistivity, and ii) the scattering with the phonons (Ref.\cite{Solyom-2007} page 389, Ref.\cite{Mott-1945} page 252) giving the main linear temperature dependence of the electrical resistivity of metals and iii) the scattering with magnons. The contribution to the resistivity due to the scattering with magnons is known to be small with respect to the other two, but has the important role to be one of the sources of the change of the spin state of the electrons in transition metal ferromagnets, i.e. the spin flip scattering \cite{Blatt-2012}. It is know from the literature that in $3d$ transition metals at temperatures well below the Curie point the spin flip events are much more rare than the momentum scattering events (with a different of at least a factor 100) \cite{Mott-1945, Coey-2010}. This fact justifies the use of the Mott's two currents model of Section \ref{The_Mott_s_two_currents_model}.

\begin{acknowledgments}
We thank Jaroslav Hamrle, Ondřej Stejskal and Milan Vrána (Charles University, Prague, Czech Republic) for inspiring discussions. This project has been funded by the Italian Ministry of University and Research (MUR) in the framework of the continuing-nature project "Next-Generation Metrology" and the Research Projects of Relevant National Interest (PRIN project Xverse T.E.C “Transverse thermoelectric energy conversion”: Grant No. 2022LLWM5F).
\end{acknowledgments}

%


\end{document}